# Economic Evaluation of the Portuguese PV and Energy Storage Residential Applications

**Ana Foles**[a,b,1], Luís Fialho[a,b,2], Manuel Collares-Pereira[a,b,3]

[a]*Renewable Energies Chair, University of Évora, 7000-651 Évora, Portugal*
[b]*Institute of Earth Sciences, University of Évora, Rua Romão Ramalho, 7000-671, Évora, Portugal*
[1]*anafoles@uevora.pt*
[2]*lafialho@uevora.pt*
[3]*collarespereira@uevora.pt*

## Abstract

In the residential sector, energy micro-generation and its intelligent management have been creating novel energy market models, considering new concepts of energy use and distribution, in which the prosumer has an active role in the energy generation and its self-consumption. The configuration of the solar photovoltaic system with a battery energy storage in Portugal is unclear in the technical, energetic and mostly in the economical point of view. The energy generation and consumption management, jointly with the battery operation, have a great influence in the configuration's profitability value. The present work evaluates different photovoltaic configurations with and without energy storage for the normal low voltage C consumer profile, for a contracted power of 3.45 kVA, to evaluate the systems' cost-effectiveness, framed in the regulation in force in Portugal - the decree-law 153/2014 -, which promotes the micro-generation and self-consumption. The analysis consists of three different geographical locations in the country, considering distinct electric tariffs. These are relevant parameters in the choice of the configuration, concluding that although the solar photovoltaic system by itself is already economical presently, its integration with battery energy storage isn't in most of the configurations, however it is already possible to find profitable PV+battery configurations, considering all the most relevant criteria, and supported by good energy management.

## Highlights

- PV-only configurations are more profitable than PV+battery in Portugal
- Mostly in Évora and Porto PV+battery configuration is becoming profitable
- In average its 22% more economic to invest in a grid-connected installation in Évora
- Best LCOE occurs with the biggest and smallest PV installations

## List of Abbreviations, Acronyms, Initials and Symbols

Analysis Period, N - the amount of time or the period an analysis covers.

B/C – Benefit-to-Cost Ratio.

Base Year - Year to which all cash flows are converted.

BTN – Normal low voltage.

BU – Battery use – quantifies the use of the battery in comparison with the sum of the energy load profile, in one year.

Cash Flow - F - Net income plus amount charged off for depreciation, depletion, amortization, and extraordinary charges to reserves.

CE – Certificate of Exploitation, needed in some of the PV configurations, defined in the Portuguese current legislation.

Contracted Power – One of the defined parameters in the electricity contract, which defines the maximum power number of household appliances which are generally used simultaneously, in the domestic sector.

DGEG – Director General of Energy and Geology.

Discount Rate - The rate used for computing present values, which reflects the fact that the value of a cash flow depends on the time in which the flow occurs.

Discount rate, d – Measure of the time value, which is the price put on the time that an investor waits for a return of an investment.

DL – Decree-Law.

DSO – Distribution System Operator.

Electricity Tariff – Price payed by the consumer for the electricity which is consumed from the electricity company, generally expressed in €/kWh.

ERSE – Regulatory Entity of Energy Services.

Feed-in tariff (FIT) – Fixed electricity prices paid to RE producers, for each unit of energy produced and injected into the electricity grid (kWh). The payment is established for the analysis period, regarding the lifetime of the project.

Inflation Rate, a - The rise in price levels caused by an increase in available currency and credit without a proportionate increase in available goods and services of equal quality. Inflation does not include real escalation. Inflation is normally expressed in terms of an annual percentage change.

Investment Year - The year in which a capital or equipment investment is fully constructed or installed and placed into service.

Investment, I - An expenditure for which returns are expected to extend beyond 1 year.

IRR – Internal Rate of Return (%).

LCOE – Levelized cost of electricity (€/kWh).

Life-Cycle Cost, LCC - The present value over the analysis period of the system resultant costs.

MiBEL – Iberian Market for Electricity.

Net-metering tariff - Incentive which allows the storing of energy in the electric grid. The surplus electricity generated by solar modules is sent to the grid, and when the producer doesn't produce enough electricity to cover its needs, purchases electricity from the grid. The balance is a credit, which is discounted in the electricity bill.

NPV – Net present value (€).

OMiP / OMiE– Portuguese/Spanish branch of MIBEL.

perspective in the base year.

PVOM - Present value of all O&M costs (€).

REN – National Electric Grid.

RES – Renewable Energy Sources.

SCR - Self-consumption rate.

SLR – Supplier of Last Resort.

SMR – Saved money rate.

SSR – Self-supply rate.

Standard Deviation - A statistical term that measures the variability of a set of observations from the mean of the distribution.

Tax Rate - The rate applied to taxable income to determine federal and state income taxes.

TLCC – Total life cycle cost (€).

TSO – Transmission System Operator.

UPAC – Self-consumption Production Unit.

UPP – Small Production Unit.

$\Delta S_n$ - Sum value of the annual cash flows net annual costs (€);

$C_{bill}$ - Electricity bill of one year, for each location and electricity tariff (€);

$C_{savings}$ – Electricity bill savings with the studied configuration (€);

$E_{Battery\ sent}$ - Energy sent to the battery (kWh);

$E_{Load}$ – Sum of the energy load profile, for one year (kWh);

$E_{supplied,m}$ - Supplied energy in kWh, in month $m$;

$F_n$ - Net cash flow, in year n;

$OMIE_m$ - average Iberian electricity gross market closing price (OMIE) for Portugal in €/kWh, in month $m$;

$PV_{consumption}$ - Energy generated through the PV system which is self-consumed (kWh);

$PV_{generation}$ - Total generated energy from the PV system (kWh);

$Q_n$ - Energy output or saved, in year n;

$R_{UPAC,m}$ - Sold energy price in €, in month $m$;

$\Delta I_n$ - Nondiscounted incremental investment costs (€);

$O\&M$ – Operating and maintenance costs (€);

# 1. Introduction

In contrast to fossil fuelled energy generation, renewable energy (RE) sources are characterized by abundancy in the environment and lower pollution. Developed countries are evolving in the sense of diversifying their energy sources, integrating micro-generation in their low voltage (LV) networks, shaping micro-grids (MG). Micro-grids could be designed for RE to fully meet the local consumption loads, considering the use of a storage unit to balance the supply and demand, considering the control of the energy flows, which can require the generation of energy or the shifting of the consumption loads in the residential load diagram. Electricity generation from RE sources can be described as dispatchable or non-dispatchable renewables, regarding the ability of the energy source to be controlled giving response to system requirements, such as the consumer loads in the residential sector. The first group includes the hydroelectric, geothermal and biomass power, and the second the wind, solar photovoltaic (PV), concentrated solar power (CSP) and wave and tidal power. RE integrated in the power grid requires changes in the existing networks, since allows bidirectional flows of energy to ensure grid stability; efficient grid management mechanisms to improve grid flexibility, response and security of supply; improvements in the interconnections (increasing capability, reliability and stability), introducing devices and methods of operation to ensure stability and control (voltage, frequency, power balance); introduction of energy storage (ES) aiming the system flexibility and security of supply [1]. The integration of solar PV modules at a residential scale allows the energy efficiency achievement, increased local reliability, reduction of energy losses, and easy architecture integration. Cost-competitiveness of solar PV and reduction of support schemes had made possible new business models to emerge, mostly in northern Europe. PV electricity generates revenues through the injection into the grid or by optimization of self-consumption, allowing the reduction of the electric bill and the growing of new energy flow models for the householder/businessman. The price decrease of solar PV modules and ES technologies will increase competition on the decentralization of energy generation/consumption models. Energy consumers are currently interested to play an active role not only in the use of RE sources, but also in the generation of RE. These consumers are referred as prosumers and are motivated by the energy bill reduction and higher control, more accurate response and environment sustainability, with clean and cheap electricity, consuming local and supporting the grid operation. Storage operation creates flexible markets, data access and management, cooperation between TSO and DSO [2].

Electric battery technologies will play a significant role in Europe’s Energy Union framework. Regarding the ten key actions designated in the SET-Plan, it is established to “become competitive in the global battery sector to drive e-mobility and ES forward” [3]. Electricity storage involves the conversion of electricity in another form of energy and is currently executed through technologies which differ in performance, characteristics and operation. ES can be conducted by pumped-hydro storage, compressed-air ES, electric batteries, superconducting magnets, flywheels, super-capacitors, chemical storage and thermal storage, or can be obtained through end-use technologies, such as plug-in electric vehicles [1]. New and cost-effective storage technologies are being developed. Apart from mitigating power fluctuations, ES systems can play other roles with PV technologies, such as load-shifting (storing energy during low demand periods and discharging in high demand periods).

Compared to other storage options, mentioned above, batteries have become popular in residential appliances due to general simplicity, materials availability, technology maturity and relatively low cost. According to BNEF, the average price of lithium-ion battery technology was 1160 $/kWh in 2010, 176 $/kWh in 2018, and for 2030 the expected value is 62 $/kWh [4]. The battery technologies and the most used correspondent applications are shown in Table 1.

Table 1 - Battery technologies and grid applications [5].

| **Application/ Technology** | **Lithium-ion** | **Lead Acid** | **Sodium-sulphur** | **Flow batteries** |
|---|---|---|---|---|
| Load shifting – reducing excess renewable energy curtailment | Suitable | Suitable | Suitable | Suitable |
| Frequency restoration reserves | Suitable | Unsuitable | Unsuitable | Unsuitable |
| Capacity reserves | Suitable | Suitable | Suitable | Potentially suitable |
| Transmission and distribution system upgrade deferral | Suitable | Suitable | Suitable | Potentially suitable |
| Voltage support | Potentially suitable | Unsuitable | Unsuitable | Unsuitable |
| Spinning reserve | Potentially suitable | Suitable | Potentially suitable | Potentially suitable |

China is the leader in PV solar energy installations, followed by USA, Japan, Germany and Italy. As market leader, China has in force exclusively photovoltaic policies as the "13$^{th}$ Solar Energy Development Five Year Plan (2016-2020)" implemented in 8$^{th}$ December 2016, in which committed to reach to 105 GW of solar photovoltaic capacity. Since 2010, a programme provides upfront subsidies for grid-connected rooftop and BIPV [6]. Spain had Royal Decree 900/2015, but currently the Royal Decree Law 244/2019 is in force, and accounts with different self-consumption schemes, defines communal self-consumption, simplifies the remuneration related with surplus energy for PV installed power no larger than 100 kW (monthly net-metering) [7]. The "National Renewable Energy Action Plan 2011-2020" defined a 20.8 % share of generated renewable energy sources in gross final energy consumption [6]. France works almost exclusively with feed-in tariffs, and it does not have a self-consumption scheme, although a community power scheme has been studied. The photovoltaic feed-in tariff is in force since 2006, last updated in 2016, and has two main variants: building installations smaller than 100 kW, and it is adjusted every semester; and tenders for buildings installations larger than 100 kW and ground-mounted plants. Targeted a 32 % of RES in gross final energy consumption, established in July 2015, 40 % in electricity and 15 % in transports. Italy has made a storage system regulation in 2015 identifying technical specifications to include storage into the national electricity, and the "National Energy Strategy" approved in 2017 promotes the integration of storage systems to accommodate the growing penetration of RE sources. The solar photovoltaic financial incentives which started in 10$^{th}$ July 2012 were cut in 25$^{th}$ June 2014. MiSE has presented provisions which will grant financial incentives to purchase electric or hybrid vehicles, or low carbon emission ones, up to the end of 2021 [8]. In 1st March of 2016, Germany has started a subsidy for solar photovoltaic installations with battery storage for residential installations: the scheme offers soft loans up to 2000 €/kW for solar photovoltaic systems and capital grant covering up to 25 % of the eligible solar panel. These values are updated (downwards) every six months. The National Energy Action Plan in force in Germany was implemented in 2010 and has the 2020 targets for 18 % of energy generated from RE, through 37 % of electricity and 13 % for transports coming from RES. United Kingdom started a feed-in tariff for renewable electricity, including solar photovoltaic, in 2010 and last updated in 2015, for small-scale (less than 5 MW). Targets in 2020 are that RES represent 15 % in gross final energy source, 31 % of electricity and 10 % of energy demand [6]. In India, The Uttar Pradesh Electricity Regulatory Commission defined net-metering regulations for rooftop solar photovoltaic, running for 25 years. The tariff is set to 7.08 INR/kWh and has entered in force on 20th of March 2015. India's Ministry of New and Renewable Energy committed to a target of 175 GW in 2022, where 100 GW comes from solar photovoltaic, 60 GW from wind and 16 GW from biomass and small hydro [6]. The Department of the Environment and Energy of Australia 2016 provides funds for community groups in regions across the country to install rooftop solar photovoltaic, solar hot water and solar-connected battery systems. The Renewable Energy Target (RET) is designed to deliver a 23.5 % share for renewables in Australia's electricity mix by 2020 [6].

The number of RE applications in Portugal is increasing according to the license requests. In March 2018 the electricity generation from RE was bigger than the effective consumption of electricity in Portuguese continent. The current Portuguese Secretary of State of Energy announced the Portugal's first dedicated auction for 1.35 GW for mid-2019 and 700 MW by 2020, tending to 50-100 MW for dispatchable renewables, claiming Portugal urgent need to move towards storage technologies. Two specific photovoltaic auctions promote the integration of PV technology from 572 MW in 2018 to 1.6 GW by 2021 and 8.1 GW to 9.9 GW by 2030 [9]. The main supplier and distributor of electricity in Portugal, EDP, is going to build the first PV plant, 3.8 MW, coupled with lead-acid batteries storage, conceived for self-consumption, in Castanheira do Ribatejo and Azambuja [10]. It becomes urgent to work further in the PV+battery market, which has lack of standards and safety rules, especially in the residential sector.

The Portuguese electrical system is divided in three main activities: generation of electricity, transmission of electricity through very high and high voltage grids, distribution of electricity through high, medium and low voltage grids and the supply of electricity to consumers. The generation of electricity is divided into ordinary regime, which corresponds to thermoelectric plants, and special regime, which includes the generation through RE sources, cogeneration and small production and generation related to other special regimes [11]. The transmission is carried out under an exclusive public service concession contract made with the Portuguese State and the REN - "Redes Energéticas Nacionais, SGPS, S.A. – (TSO). TSO must connect all the entities to its network if the connection is feasible technically and economically, and if the applicant satisfies the requirements for connection. Regarding supply, there are two regimes:

1. Free market supply to eligible consumers – Eligible consumers since 4th September of 2006; the supply is made by free marketers using freely negotiated conditions (except some ERSE's Regulation terms);
2. Supplier of last resort (SLR) – This supplier has a supply license and must ensure specific consumers with regulated tariffs (ruled annually by ERSE). This supplier must buy all the special regime generation at fixed and regulated prices depending of the generation technology (under feed-in tariffs scheme). This doesn't prevent the SLR generators to sell their energy to other suppliers.

In a free market regime, the participants involved in the production can sell the produced electricity and the ones who need electricity can buy it, whatever the finality.

Portugal and Spain have been integrating their electricity markets into one, the MIBEL (Iberian Electricity Market). They have a shared spot market operator, the OMIE, which operates since July 2007, and a forward market operator, the OMIP, since July 2006. The MIBEL market is based in a group of contracting modalities which complement each other. The OMIE spot market is regulated by Spanish legislation and OMIP by Portuguese legislation (under the MIBEL International Agreement), being acknowledged by the legislation of the other country. ERSE establishes regulations, and of these the most important regulations to consider are about commercial relations, tariffs, quality of service, access to networks and interconnections and networks operation. DGEG and independent regulatory entities are responsible by the regulation enforcement. Their responsibilities are issuing, amending and withdrawing licenses for electricity generation, maintaining registries of electricity supply, and supervising the security of supply. REN owns and maintains on an exclusive basis the electricity transmission system in the Portuguese continent. The distribution system operator of the high and medium voltage is the EDP – Distribution SA and has the concession of most low voltage municipal distribution systems. In Azores the distribution operator is "Eletricidade dos Açores" (EDA), and in Madeira is "Empresa de Electricidade da Madeira" (EEM). Supply is carried out by several companies, the main supplier of last resort is EDP Serviço Universal in the continent, and in Azores and Madeira are the same as mentioned for distribution.

The electricity produced by Portugal is enough to meet the consumption needs, but for commercial reasons, Portugal imports electricity from Spain. In 2017 Portugal imported 3,072 GWh. The surplus production from Portugal is exported to Spain. The electricity generation from RE sources has contributed significantly to an exporter balance, in the last few years.

Natural gas and coal are the main fossil sources of energy generation in Portugal, nuclear does not exist and the RE production has increased in the last few years. [11]. In the past there were support mechanisms for RE technologies based on feed-in-tariff system, tax benefits and investment subsidies. Currently there are no support mechanisms, except for offshore wind and wave energy (new technologies) and small cogeneration. The Directive 2009/28/EC promotes the energy from RE, setting a target of 20 % share of final energy consumption in 2020 (PNAER 2020), and 10 % of transport fuels coming from renewable sources by 2020. On 30 November 2016, European Commission published a proposal for a revised RE Directive to make the EU a global leader in RE and ensure that the target of at least 27 % renewables in the final energy consumption in the EU by 2030 is met. Recently, European Union (EU) has settled an at least 32 % share of final energy consumption in 2030 as global leader.

In 2015 Portugal has made a strategic plan, the "Green Growth Commitment 2030", which quantifies the targets for 2030, namely the 31 % of RES in gross final energy consumption by 2020 and 40 % by 2030, besides many measures to increase sustainability and energy efficiency. Portugal 2013-2016 energy plan was set by the PNAEE 2016 document (Plan of Action for Energy Efficiency) and currently is set by PNAER 2020 (National Plan of Action for Renewable Energies 2013-2020, approved by Ministers' Council Resolution No. 20/2013 of April 10. Portuguese Government has committed internationally to reduce its greenhouse gases emissions to achieve carbon neutrality by 2050. This has risen as a form of report as "Roteiro para a Neutralidade Carbónica" – Carbon Neutrality Road Map. Besides Portugal being a small country, the other main obstacle to develop renewable energy is the lack of interconnections of the Iberian Peninsula (Portugal and Spain) and other European Countries or the North of Africa.

In Portugal 2013 (Order from the Director of Energy and Geology from 2013, December 26) a feed-in-tariff scheme was implemented for micro and mini generation, which is currently superseded by the Law on Self-consumption Decree-Law 153/2014 [12], in force since the 20th October of that year. This Decree-Law establishes the legal regimes of the RE self-consumption, considering two types of units – the UPP (Small Production Unit) – which includes the former micro and mini generation systems up to 250 kWp, and where the electricity production is exclusively sold to the grid operator, and the installation consumption is exclusively supplied by RESP - and the UPAC (Self-Consumption Production Unit) – which considers the self-consumption based on renewable technologies, making possible to sell to the grid the surplus energy generation. When the generated energy by UPAC meets the demand, the produced energy supplies the consumption point, when the generated energy isn't enough, the RESP supplies the consumption point. This decree law defines some licenses, installation audits and paying regimes of the electricity sold to the grid. DL 153/2014 has established a distribution generation model, which promotes the decentralization - generation of energy near the consumption point -, the generation of energy by RE, the increase of the competition and the security in supply, the reduction in peak power requirements, the encouragement of the PV industry growing as well as the communities. Regarding the two decentralized electric energy generation in force in Portugal, the UPP is based on a single generation technology and total injection in RESP. The UPAC, as PV RE, allows the connection of the system with the grid (RESP), giving priority of its consumption in the installation, still having the opportunity of selling the surplus, when applicable. The possibility of connecting the UPAC to the grid should be studied, to conclude its profitability and decide whether to connect it or not. Because this regime is the object of study of this work, the general regulation for UPAC will be detailed, which establishes:

- Connection power being less or equal to 100 % of the contracted power of the consumer installation;
- The generated electricity from UPAC should be near of the consumption point in the installation;
- If it is connected to the electric grid, the instantaneous generation surplus could be sold to SLR;
- The consumer can install an UPAC for each electric installation, consume the generated energy or export its surplus to the grid. The Production Unit (UP) is installed in the same site of consumption. The consumer could have multiple registered UPs, although each installation is associated with a single UP.
- If a 1.5 kW UPAC is connected to the electric grid, the consumer is obliged to have a dedicated electricity metering equipment, to account the injected electricity.

The contract should be concluded with a maximum term of 10 years, renewed for periods of 5 years. If the UPAC installed power is higher than 1.5 kW and is connected to RESP, the consumer has a monthly fixed compensation for the first 10 years after receiving the certificate of exploitation (CE). The licensing process is made through electronic register in the SERUP site (UPs register), managed by the DGEG authority, submitted by the proprietary of the installation, and its summary is given in Figure 1. Table 2 presents the fee charges of the DL 153/2014 of the UPP and UPAC regimes.

Figure 1 – Resume of DL 153/2015 regimes, the UPP and the UPAC.

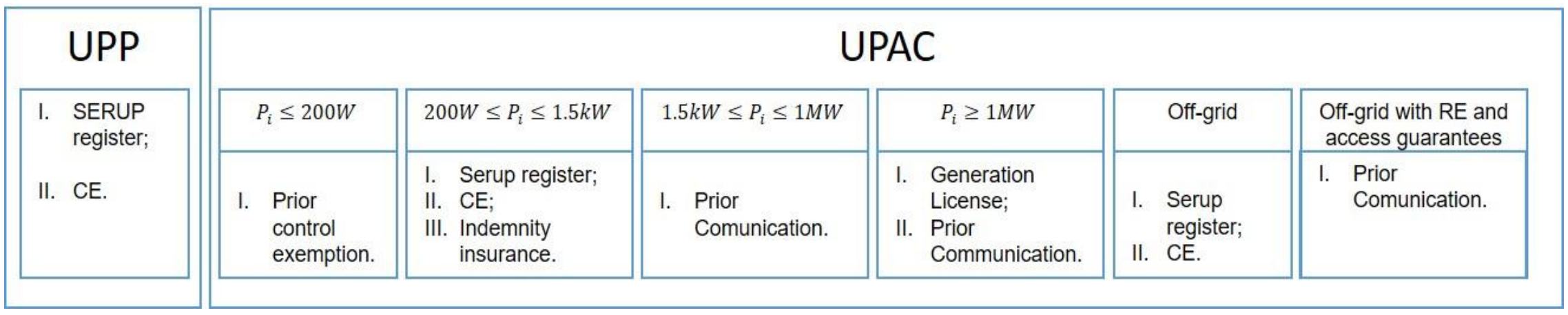

Table 2 - Fee charges applicable to UPAC regime with and without grid injection, regarding the installed power in kW – DGEG (Portaria 14/2015) remuneration*.

| Installed Power Capacity | UPAC charges with grid injection | UPAC charges without grid injection |
|---|---|---|
| < 1.5 kW | 30 € | N/A |
| 1.5 kW – 5.0 kW | 100 € | 70 € |
| 5 kW – 100 kW | 250 € | 175 € |
| 100 kW – 250 kW | 500 € | 300 € |
| 250 kW – 1000 kW | 750 € | 500 € |

*These fees are not currently charged; its end or great reduction is expected in future versions of the Portuguese legislation given the approved European Union Directive (RED II)[1], providing the exemption of fees and charges for small self-consumption facilities (up to 30 kW) and the possibility for communities to generate, store and sell the surplus generation.

Through the consulted works, regarding this topic, [13] present a techno-economic study based in future price scenario which considers the application of PV and battery energy storage in the Azores island, with three battery sizing for each battery technology, the lithium-ion and vanadium redox flow. The aim is the minimization of the cost of electricity generation, and the used economic indicators are the NPV and ROI. In [14] an economic analysis is made considering lithium-ion and lead-acid battery technologies with different RE sources applied in India with net-metering, addressing the advantages of integrating energy storage in the networks, and considering real load and resource profiles data and component prices, concluding that lithium-ion batteries are more viable to apply in those cases. In [15] an analysis was made for Almeria, Spain, and Lindenberg, Germany, assessing impacts of orientation and tilt angles in the self-consumption, with storage. Higher load profiles showed better results in self-consumption, trade-off in self-consumption increase and cost reduction of investment, in the residential sector, and framework regulations. Applied in Australia, [16] study the PV+battery configuration, using NPV, IRR and LCOE in application in Australia, concluding that PV only systems are profitable, instead of the PV+battery, and that the economic losses of adding a battery can only balance the benefits that it brings to the grid. In [17] a residential analysis is made for three USA locations, with the configuration PV+lithium-ion battery, concluding that can compete with grid prices with adequate sizing in those locations, using the LCOE indicator. In [18] mono-crystalline PV systems cases and three lead-acid battery cases are studied, to be applied in Italy, without subsidies, considering also lead-acid batteries. Relevance of the discounted cash-flows (DCF) is highlighted, jointly with NPV, giving relevance to the variables of PV and electricity, associated costs, profiles and batteries. In [19] five different cases of storage with net-metering are studied. Main indicators are for three locations in Italy, PV and battery sizing and installation costs. It concludes that economic feasibility is far, and losses generated by storage are a disadvantage. Regarding the Portuguese context, some relevant approaches have been made considering current legislation, such as the work of [20] which carries a complete economic analysis using the NPV, LCOE, BCR and IRR as economic indicators to evaluate four configurations of PV and OPzV gel batteries (lead-acid), on a 25 year lifetime analysis, using PV kits, and concluding that most of the configurations weren't economical. In [21] economic indicators DPB and IRR are used, and legislation in Portugal is clarified. An analysis is conducted for different sectors and three different locations (Lisbon, Porto and Faro), with PV systems with different azimuth and tilt angles, evaluating self-consumption, remarking the importance of the tariff, load profile and PV surplus generation. In [22] PV+battery impact is studied, considering two storage control strategies and tariff fee charges, showing that all the configurations are profitable with a payback below 10 years.

[1] Directive (EU) 2018/2001 of the European Parliament and of the Council of 11 December 2018 on the promotion of the use of energy from renewable sources.

From the literature revision made, various studies have been conducted to estimate optimal PV+battery configurations, based in the most used economic indicators, for application in the Portuguese context, for the residential case. Recognising the important work done, the present work stands out in the way of evaluating the PV configurations in three different locations in Portugal, for two electric energy tariffs, with current justified market prices. Detailed Portuguese electricity sector remarks are given and legislation for energy micro-generation is clarified, to contribute to a better integration of PV-only and PV+battery configurations in the residential sector in Portugal.

Present work is structured as follows. The second section aims to clarify the used methodology for the economical and energetic assessment. Third section presents the case study justification and explanation context. Fourth section presents the results of this work. Section five shows the main conclusions of this research.

## 1.1. UPAC Characterization

As explained, the licensing scheme of most installations includes the production unit registry in the SERUP database [23], which has made available some data about the concluded and rejected RE installations. Regarding PV installations in the UPAC and prior communication (MCP) regimes, the data made available is from March to December of 2015, January to December of 2016 and from January to July of 2017. The correspondent data comprises 1843 UPAC installations with 95,995 MW and 12363 MCP installations with 10,845 kW, comprising a total of 106.8 MW of installed power in Portugal, and can be observed in Figure 2 and Figure 3.

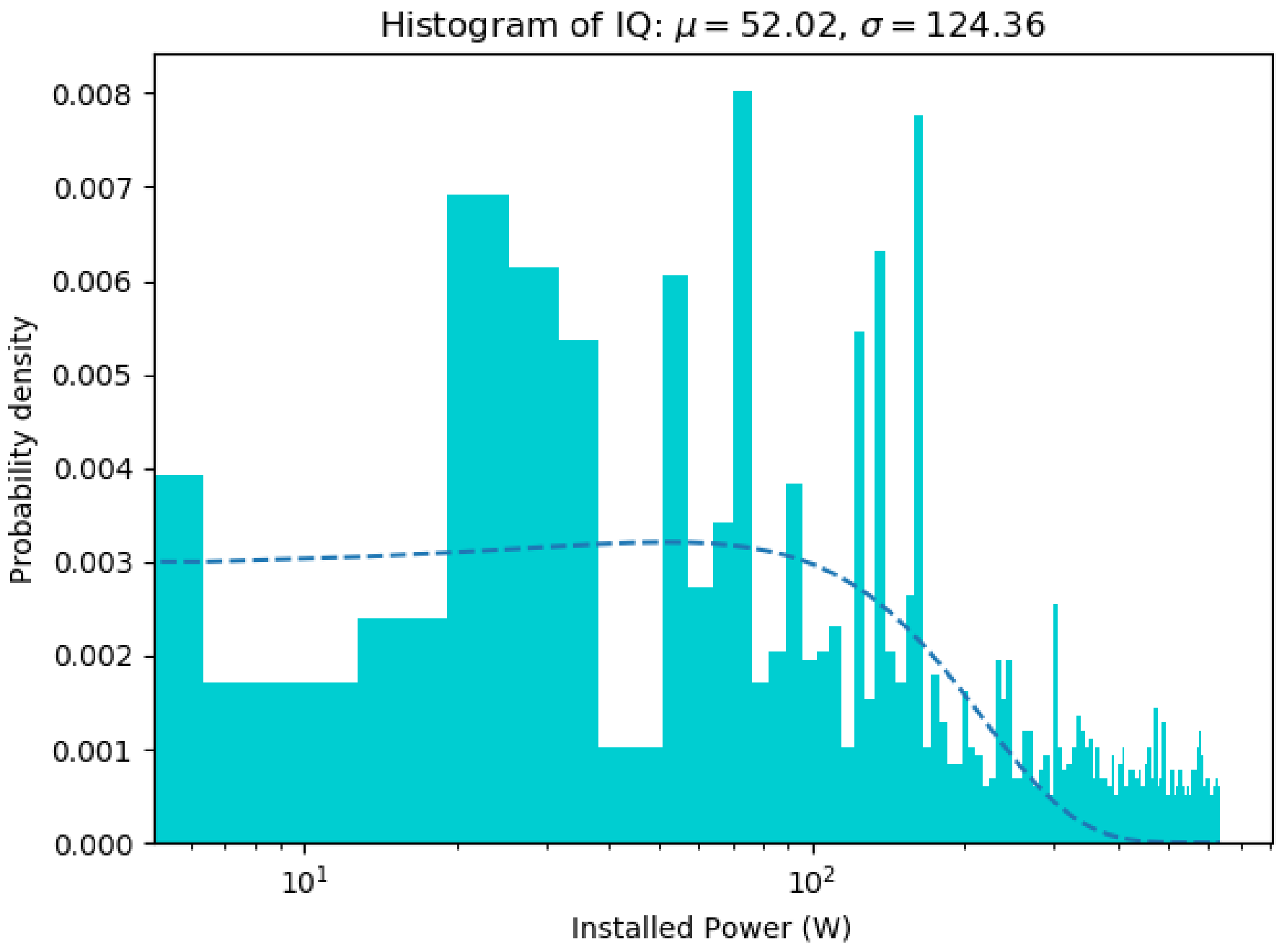

Figure 2 - Installed power distribution of the Portuguese UPAC registered installations from the available data of the SERUP database [23], with 100 bins.

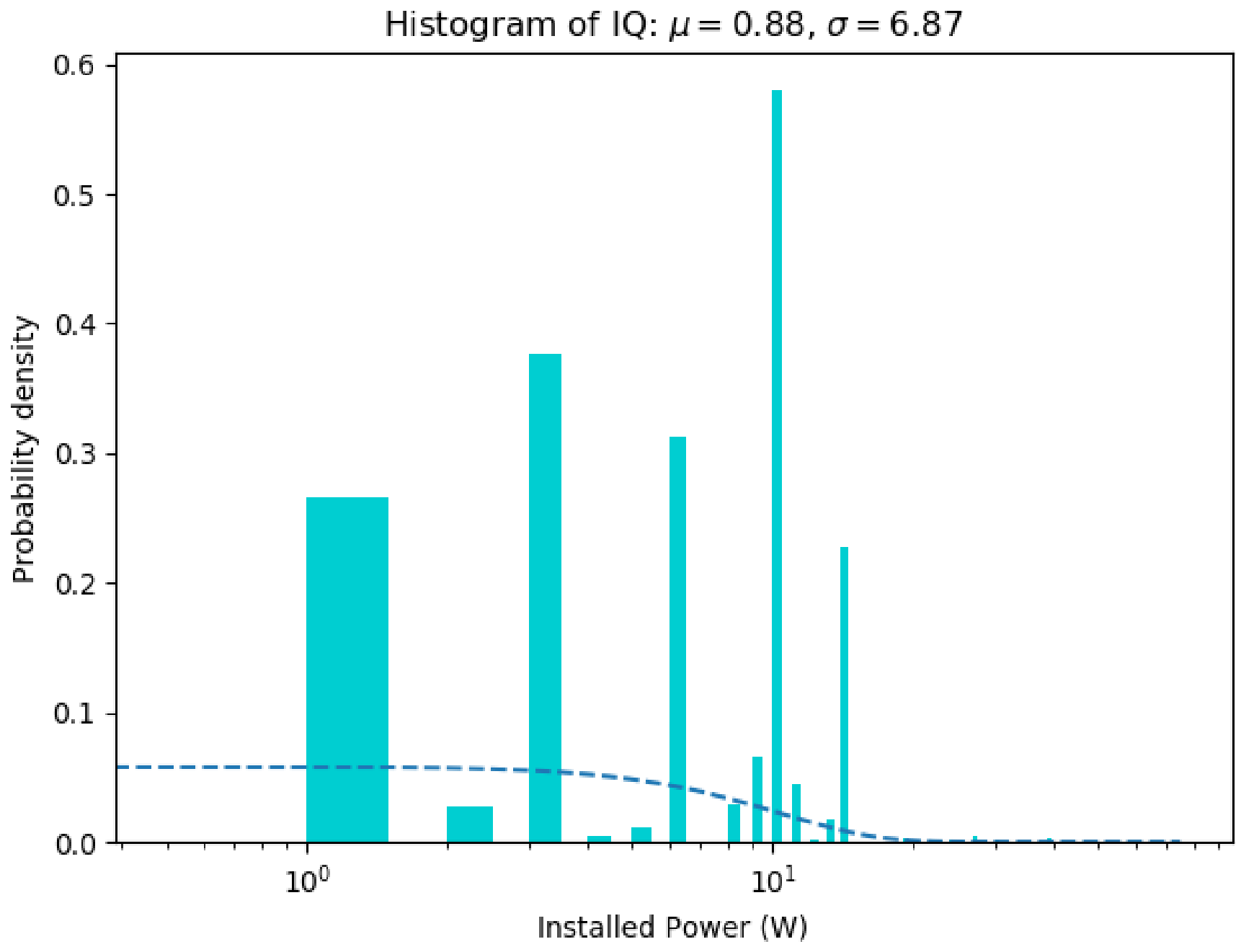

Figure 3 - Probability density of the logarithm MCP installed power (W), made available from the SERUP database [23], with 150 bins.

A domestic costumer with a suitable sized photovoltaic system in the UPAC regime produces energy and can use it to exclusively supply his loads, generally called exclusive self-consumption. Since the produced energy by the PV system is variable through the day, seasons and years, usually for the household consumption the electricity generated by the photovoltaic system has a surplus or isn't enough to totally satisfy the domestic loads. In the first case, the energy surplus can be curtailed, injected into the grid, or stored in batteries for later consumption. If the generated electricity isn't enough to totally supply the loads, the resultant consumption needs must be supplied from the electric grid or from other energy source.

# 2. Methodology

A complete technology or project investment assessment is an annual investment analysis, considering all the relevant costs, revenues, taxes and rates. The objective of making an economic analysis is the provision of relevant information to make a judgement or a decision [24]. Present work based the analysis in a 25 years lifetime. The economic figures used in this work are the Net Present Value (NPV), the Total Life Cycle Cost (TLCC), the Levelized Cost of Energy (LCOE), the Simple Payback Period (SPB), the Internal Rate of Return (IRR) and the Benefit-to-Cost Ratio (B/C ratio). Although the importance of the economic assessment in the decision making, using energy indicators is also crucial. Despite these last parameters being used in smaller time periods (for example in the analysis of a single energy strategy), we decided to calculate some energy parameters for one year, for one year of generation and consumption. The evaluated key performance indicators, only physically tangible after the payback time, are the self-consumption ratio, the self-supply ratio, the battery use, and lastly a grid independency rate based on the achieved savings.

## 2.1. Measures

The NPV examines the cash flows associated with a project, over the duration of the project. It is the value in the base year (usually the present), and can be expressed as follows in Eq. (1),

$$NPV = \sum_{n=0}^{N} F_n / (1+d)^n \tag{1}$$

where $F_n$ is the net cash flow, in year $n$; $N$ is the period of the project; $d$ is the annual discount rate. This parameter is generally recommended to evaluate the characteristics and decisions of the investment, and social costs. The NPV value can have some variations, as the calculus includes or not the after-taxes values and being in current or constant euros. If the NPV value is positive the project is considered economical and can be accepted; in contrast, if the NPV value is negative, the project isn't economical, meaning that returns are worth less than the initial investment, being an indicator of a no-good decision. In theory, if the NPV value is null, the investor should be indifferent whether to accept the project or not. The applicability of this indicator should be carefully analysed, since these considerations aren't valid for all applications.

The TLCC evaluates differences in costs and the timing of costs, between alternative projects. These costs are referred to the asset acquisition, costs in its life cycle or in the period of interest to the investor. Only the relevant costs are considered, and are discounted to a base year, recurring to the present value analysis. TLCC has three variations, considering no taxes, after tax deductions or before-tax revenue required. In this work, the more suitable form is the no taxes formula, adequate to residential/non-profit/government application, expressed in Eq. (2),

$$TLCC = I + PVOM \tag{2}$$

where $I$ is the initial investment, and $PVOM$ is the present value of all O&M costs, as can be seen in Eq. (3).

$$PVOM = \sum_{n=1}^{N} O\&M_n / (1 + d)^n \tag{3}$$

The LCOE is the cost of each unit of energy produced or saved by the system, over the period of the analysis, which will equal the TLCC, discounted back to the base year. In other words, it could be explained as the cost of one unit of energy, which is kept constant in the analysis period, that provide the same net present revenue as the NPV cost of the system. The levelized cost of energy is very useful to compare different scales of operation, investments and/or operating periods. There are many ways of calculating this parameter and the Eq. (4) will be used,

$$LCOE = TLCC / \left\{ \sum_{n=1}^{N} [Q_n / (1 + d)^n] \right\} \tag{4}$$

where $Q_n$ is the energy output or saved in year $n$. In the case of PV+battery configurations, the analysis was carried out considering the $Q_n$ value as the PV generation (energy output) plus the energy sent to the storage unit and effectively used (saved), in the year $n$.

The Internal Rate of Return (IRR) the rate at which the NPV of the future cash flow is set to zero. When applied, this rate brings the expenses values to the present, and make them equal to the return of the investment values. IRR obtained value is generally compared with a "hurdle rate". It allows the comparison between many different investment activities. The rate is given by Eq. (5).

$$IRR = \sum_{n=0}^{N} [F_n/(1+d)^n] = 0 = NPV \quad (5)$$

The SPB is a fast and simple way to compare investments. It is defined as the time (number of years) required for the net revenues associated with an investment of a certain project to be recovered, without accounting the time value of money. It could be described as Eq. (6) explicit.

$$\sum_n \Delta I_n \leq \sum_n \Delta S_n \quad (6)$$

Where $\Delta I_n$ are the nondiscounted incremental investment costs (including incremental finance charges), and $\Delta S_n$ is the sum value of the annual cash flows net annual costs. One of the main disadvantages of using this parameter is the fact that it ignores the value of the money over the period, which implies that the investor doesn't have opportunity cost. It also ignores the returns after the payback year. On the other way, it is simple of calculate, implement and explain.

The Benefit-to-Cost Ratio (B/C ratio) ratio of the SUM of all discounted benefits accrued from an investment to the sum of dl associated discounted costs. It is used to discover at which level the benefits of a project exceed the costs. This indicator is generally used from a social perspective. It can be described in Eq. (7).

$$B/C = [PV(All\ benefits)]/\ [PV(All\ costs)] \quad (7)$$

Where $[PV(All\ benefits)]$ is the present value of all positive cash flows, and the $[PV(All\ costs)]$ is the present value of all negative cash flows.

Energy indicators are studied for one complete year. The self-consumption rate (SCR) is a way of quantifying how much energy is generated and self-consumed locally. The SCR is generally given through the formula given by Eq. (8),

$$SCR = PV_{consumption}/PV_{generation} \quad (8)$$

where, $PV_{consumption}$ is the energy generated through the PV system which is self-consumed, and $PV_{generation}$ is the total generated energy from the PV system. The generated solar photovoltaic energy which is consumed is obtained through the subtraction of the "curtailment" losses or injected into the grid.

The self-supply rate (SSR) is an energetic indicator which quantifies the degree of autonomy from the grid, and is given by the following formula, given by Eq. (9),

$$SSR = PV_{consumption}/E_{Load} \quad (9)$$

where, $E_{Load}$ is the energy load profile.

The battery use (BU) can be described as a way of quantifying the usage of the battery, comparing the energy charged to the battery, and the energy load. This indicator can be given by the following expression, presented by Eq. (10),

$$BU = E_{Battery\ sent}/E_{Load} \quad (10)$$

where, $E_{Battery\ sent}$ is the energy sent to the battery, in kWh.

The saved money rate (SMR) quantifies the degree of autonomy from the grid in €. This indicator only makes sense to be calculated after the payback time break-even is achieved. The energy that was satisfied by the grid before the PV installation and now isn't – through self-consumption, charging/discharging the battery and injection into the grid – is quantified as money saved, as the Eq. (11) shows,

$$SMR = C_{savings}/C_{bill} \quad (11)$$

where $c_{savings}$, in €, is the money payed with the studied configuration, comparing with the $C_{bill}$, in €, which is the current electricity bill (only grid consumption), for each location and electricity tariff, for one year.

## 3. Case Study Definition

This work has the aim of comparing different photovoltaic configurations, evaluating its economic feasibility in a variety of options. The analysis is made for the Portuguese residential figure characterized through the profile consumption BTN C and for a contracted power of 3.45 kVA. Four PV power installations are studied, namely 0.50 kWp, 0.75 kWp, 1.50 kWp and 3.45 kWp, for off-grid or grid-connected connection, for three different Portuguese locations – Évora, Porto and in one island of the Azores archipelago. The two chosen continental sites represent approximately the Portuguese regions with regard the solar resource potential and the Azores as an extreme Portuguese location, 1600 km West of Portugal, in the Atlantic Ocean. The chosen power sizing for the studied cases was considered the most relevant for present work regarding the current legislation in Portugal, the DL 153/2014. For each of these PV installations sizes, two electricity tariffs were addressed - the flat tariff and the bi-hourly tariff – considering the different tariffs in the continent and Azores islands. In the sense of the residential sector, three different storage capacities were researched, namely 3.3 kWh, 6.6 kWh and 9.9 kWh for each of the studied PV sizes. In the next lines, a more detailed explanation is given.

## 3.1. Legislation – detailed case

For the case in which the generated energy by the UPAC is not fully self-consumed and is injected into the public grid (RESP), the price of the electricity sold to the RESP is given by 90 % of the average Iberian electricity market closing price, and can be expressed through Eq. (12),

$$R_{UPAC,m} = E_{supplied,m} \times OMIE_m \times 0.9 \quad (12)$$

where $R_{UPAC,m}$ is the sold energy price in €, in month $m$; the $E_{supplied,m}$ is the supplied energy in kWh, in month $m$; and the $OMIE_m$ is the average Iberian electricity gross market closing price (OMIE) for Portugal in €/kWh, in month $m$. The average monthly wholesale electricity prices for the year of 2018 are presented next, in Figure 4, data made available by OMIE [25].

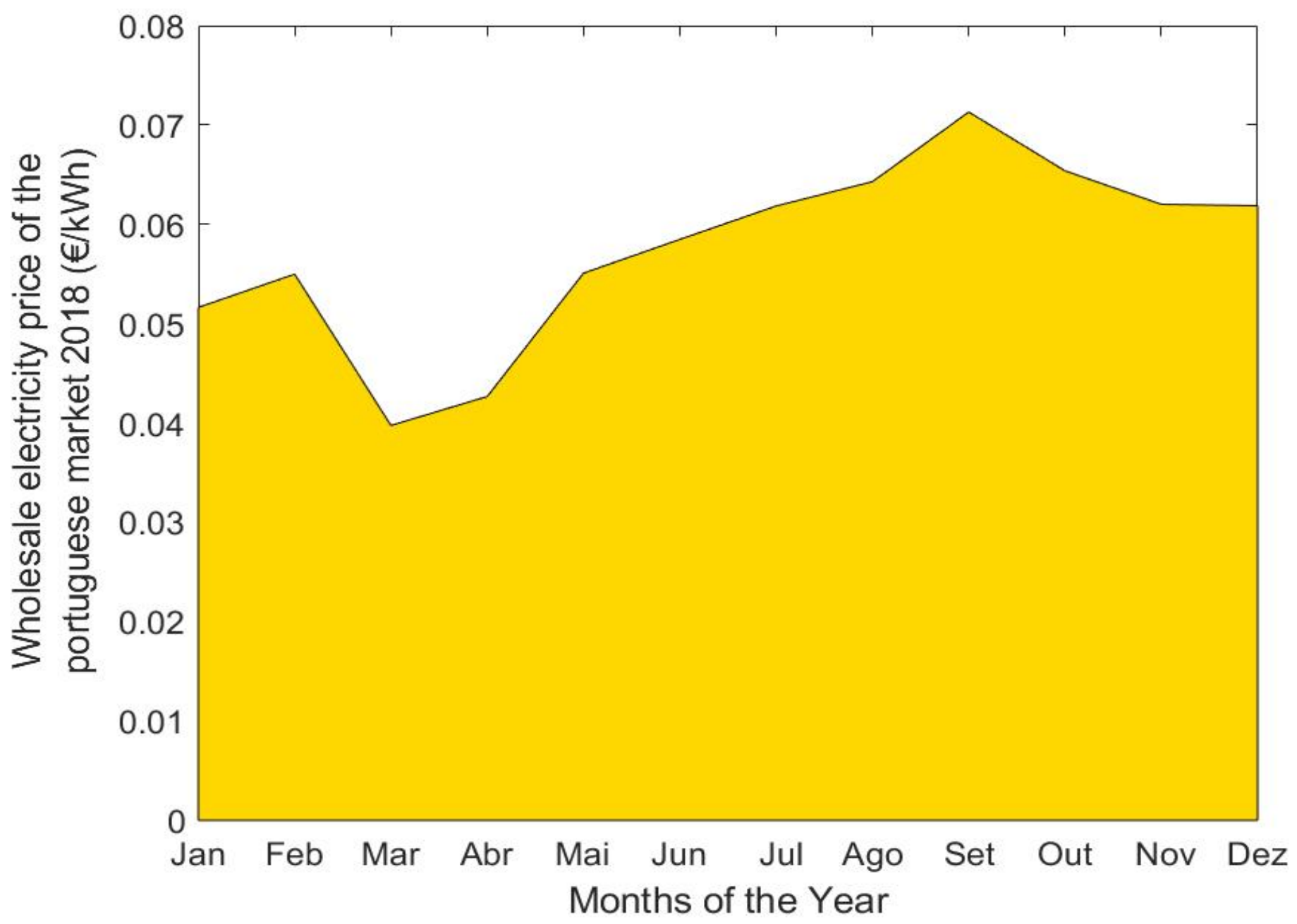

Figure 4 - Average monthly wholesale electricity prices of the year of 2018 for Portugal.

For the studied cases where the electricity is injected in RESP, the average monthly electricity prices showed in Figure 4 are used.

## 3.2. Domestic Consumption Profile

The economic analysis was performed using the average consumption profiles of normal low voltage (BTN) provided by EDP Distribuição [26]. The company collected consumption data from each fifteen minutes and organized it by sector of activity (A, B and C), which made possible the estimation of the electric consumption for the year of 2019, based in the previous year's Portuguese continent consumption. Residential sector - BTN C (contracted power less than or equal to 13.8 kVA and annual consumption less than or equal to 7140 kWh) -, data used in this study is shown below in Figure 5.

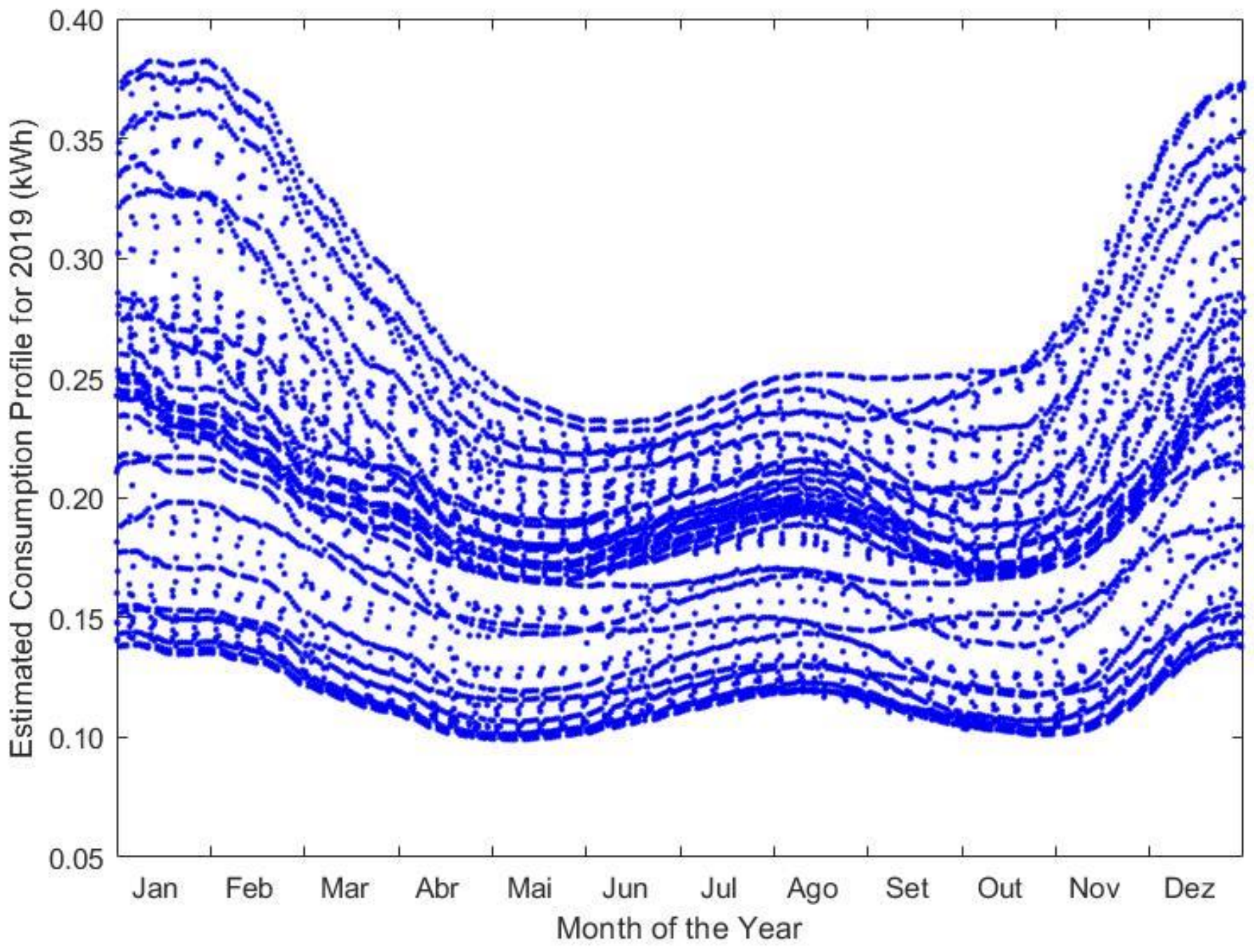

Figure 5 – Estimated electric domestic consumption profile of BTN C for 2019, made available by EDP Distribuição [26].

## 3.3. Solar Photovoltaic Irradiation Profile

Three locations were chosen to study the application of PV-only and PV+battery configurations: Évora, Porto and Azores island, and its geography can be seen in the map of Portugal in Figure 6. Évora is the capital of Alentejo, a region in the centre-south of Portugal, a city characterized by an average annual sum value of global horizontal irradiation (GHI) of 1846 kWh/$m^2$ [27], defined as one of the best locations regarding solar irradiation availability in the South of Europe. Porto is the second biggest city of the Portuguese continent and is described as the city that was in the origin of Portugal. As a coastal region in the north of Portugal, has interest of study through its average annual GHI near of 1706 kWh/$m^2$ [27]. Azores is a Portuguese archipelago with nine islands, and although being the region of these three with the lowest annual GHI, near 1307 kWh/$m^2$ [27], it is a region with lack of registered residential solar PV installations as can be observed in Table 3, and is relevant to evaluate a case with different electricity tariffs from the continent. For each of the sites, a simulation in SISIFO [28] online simulator was carried out, considering south orientation and optimum inclination for each region, and the hourly irradiation for each month of the year was extracted. The average values were considered and used in the present simulation.

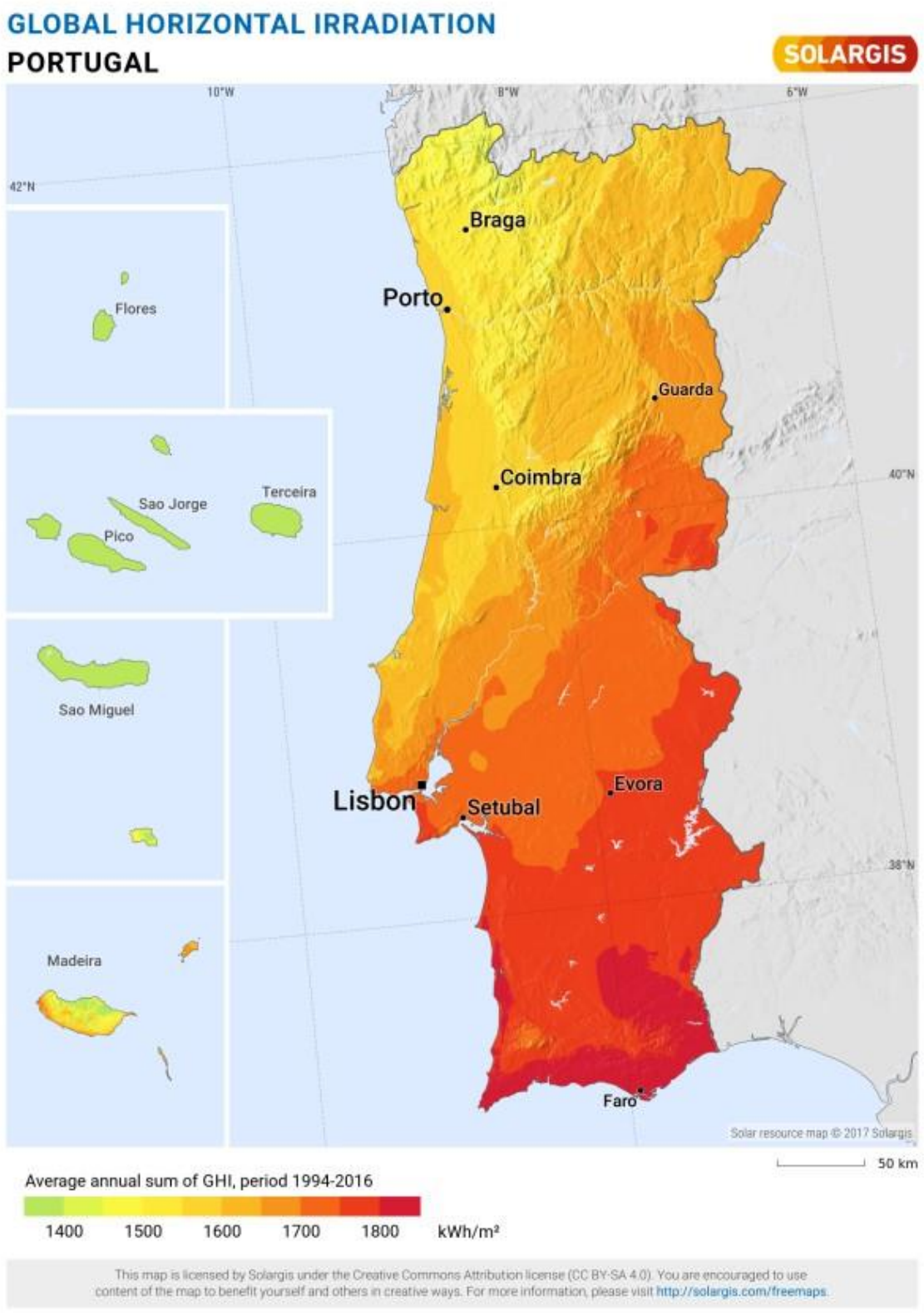

Figure 6 - Solar global horizontal irradiation (GHI) map of Portugal territory [29]

Table 3 - Electricity generation balance in the Azores, from January to December 2018 [30].

| Source | Azores | Continent |
|---|---|---|
| Fossil | 61 % | 47 % |
| Geothermal | 26 % | 0 % |
| Hydroelectric | 3 % | 24 % |
| Wind | 8 % | 22.5 % |
| RSU | 2 % | 0 % |
| Bio | 0 % | 5 % |
| Solar | 0 % | 1.5 % |

## 3.4. Electricity residential tariffs

The electricity price in Portugal is structured with three tariff regimes, namely the flat, bi-hourly and tri-hourly tariffs. Flat tariff means the load consumption is equal for every hour of the day, indicating that the tariff is independent from the period of consumption. Bi-hourly and tri-hourly tariffs distinguishes, respectively, two and three periods of consumption, by attributing two or three electric tariffs, for off-peak and peak hours. For the bi-hourly and tri-hourly tariffs, two main variants exist, the daily cycles and the weekly cycles, and these differ in the time of the year. The bi-hourly tariff considers the weekly cycle tariff, which has variations in summer and winter legal times, and the daily cycle tariff is constant through all the year. The tri-hourly considers the weekly cycle, with variations on the summer and winter legal times, so as the daily cycle. In this work, we choose to work with daily cycles in the low voltage network. For the locations of Évora and Porto, located in the Portuguese continent, the used tariff is from EDP Comercial company [31] and for the Azores island is used the EDA tariff [32], [33].

Figure 7 presents the daily cycles used in this work.

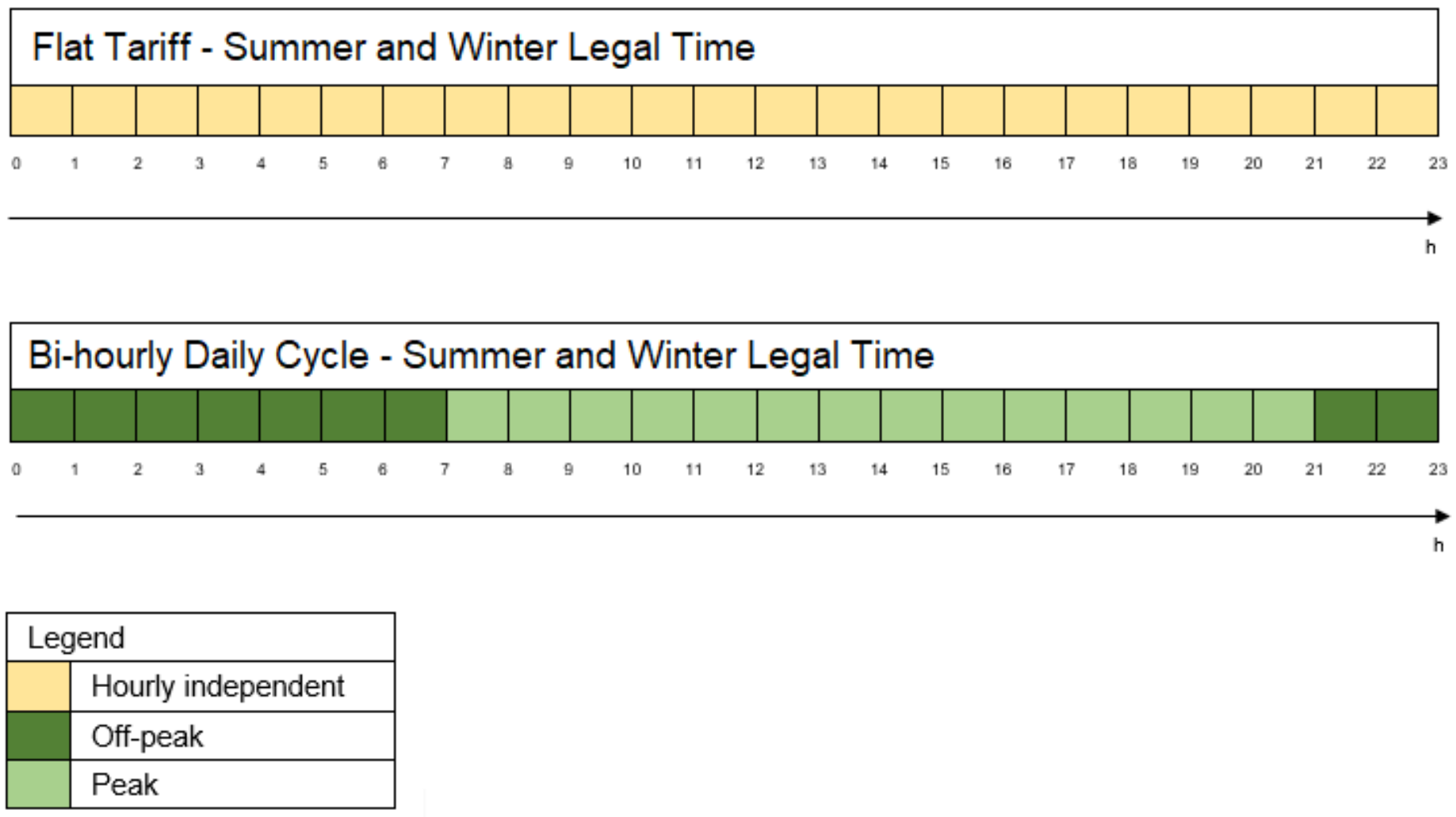

Figure 7 - Peak and off-peak periods of the daily cycles, indifferently for summer and winter legal times [34].

Regarding the stored energy, it is valuated at the valid price on the moment of its use (usually in a later period of the day), depending on the consumer contract and on the time at which the consumption occurs. Generally, for bi and tri-hourly tariffs the peak periods are associated with higher tariffs, conversely of the off-peak ones. In the storage sizing, special care must be given to the number of charge and discharge cycles over the lifetime of the battery, and the difference of the output of the chosen tariff. Concerning the tariffs used in this work, Table 4 presents the electricity tariffs for the continent and island used in this work.

It should be noticed that since the work addresses the daily cycles, when working with a bi-hourly tariff, the legal time from summer to winter or vice-versa don't have influence in the results, since it is off-peak time in the time-shift hour.

Table 4 - Flat and bi-hourly tariffs of EDP Comercial of the Portuguese continent and from EDA in the Azores, for the year of 2019.

| *Tariff* | *Flat* | | *Bi-hourly* | | | |
|---|---|---|---|---|---|---|
| *Location* | *Évora/Porto* | *Azores* | *Évora/Porto* | | *Azores* | |
| *Regime* | Normal | Normal | Peak | Off-peak | Peak | Off-peak |
| *Contracted Power (€/day)* | 0.2187 | 0.1648 | 0.2282 | 0.2282 | 0.1694 | 0.1694 |
| *Energy (€/kWh)* | 0.1493 | 0.1607 | 0.1867 | 0.1098 | 0.1908 | 0.1000 |

As a simplification and in order to be able to assign a value to the electricity in off-grid systems, it was considered for these installations the same energy cost and tariff structure as those connected to the grid.

## 3.5. Energy Storage and Inverters

The studied cases consisted of three lithium-ion battery capacities, namely B1 with 3.0 kW/3.3 kWh, B2 with 3.0 kW/6.6 kWh and B3 with 3.0 kW/9.9 kWh. Residential energy storage market has increased in recent years, given place for new, more efficiently and more environmentally friendly technologies to arise. An example of this are the lithium-ion batteries, which adequately fit in this application and have good cycle efficiency. In this work, we chose to work with market available lithium-ion batteries, mainly due to the trade-off of efficiency, energy density and current market price.

Regarding the battery specifications, special attention must be given to its capacity lifetime degradation, depth of discharge and lifetime. To calculate the energy that is stored in the battery storage system and is then effectively used by the prosumer, some aspects must be considered, as the power electronics efficiency, the battery efficiency (charge and discharge) considering the yearly battery degradation, and finally the depth of discharge. The chosen battery characteristics given by the manufacturer are shown in Table 5.

For the smallest PV configurations, the batteries capacities are unfit, comparing with the biggest PV power installations. Although this consideration, the smallest PV power installations with storage are seen not as a following case, but only as comparison cases with the biggest PV installation ones.

Table 5 - Lithium-ion battery characterization data, given by the manufacturer [35].

| Battery Identification | B1 (3.0 kW/3.3 kWh) |
|---|---|
| Model | METERBOOST-48-LTO6-3.3 |
| Nominal Voltage (V) | 48.0 |
| Maximum/minimum Voltage (V) | 32.0-58.4 |
| Nominal Capacity (Ah) | 63 |
| Nominal Capacity (kWh) | 3.3 |
| Nominal Power (kW) | 3.0 |
| Weight (Kg) | 17 |
| Length x Width x Height (mm) | 430x360x76 |
| Useful life-cycle (years) | >17 |

Micro-inverters were chosen in some cases, one with 0.50 kW of nominal power: the APS YC 500 micro-inverter; and one with 0.25 kW of nominal power: the APS 250. The hybrid inverter is the Solax SK-SU3000E X-HYBRID SERIES G2.

### 3.6. Proposed Scenarios and Economic Considerations

The present work has studied four main scenarios, summarized in Table 6. All the different configurations are simulated for each scenario.

Table 6 - Summary of the proposed scenarios.

| Case | *Photovoltaic System* | | *Grid Consumption* | | *Storage* | | *Surplus Electricity (Priority)* | | |
|---|---|---|---|---|---|---|---|---|---|
| | *Off-grid* | *Grid-connected* | *Yes* | *No* | *Yes* | *No* | *Battery* | *Grid* | *Waste* |
| I | x | | | x | | x | | | x |
| II | | x | x | | | x | | x | |
| III | x | | | x | x | | x | | |
| IV | | x | x | | x | | x | | |

- Case I – The domestic prosumer has a photovoltaic system used to perform exclusive self-consumption. The surplus of the solar photovoltaic generated electricity is wasted. For the periods in which the photovoltaic generation is not enough to supply the consumer’s load diagram, being off-grid the consumer will be without power supply (at night) and/or should do a careful load management.
- Case II – In this scenario, the prosumer’s solar photovoltaic system is grid-connected, and self-consumption is used. The surplus of the generated electricity is sold to the grid. In the periods when the generation isn’t enough to supply the loads, the prosumer consumes electricity from the grid.
- Case III – The domestic consumer has a photovoltaic system which performs self-consumption. The electricity surplus is stored in the battery, the storage use is a priority. If the battery reaches its maximum state of charge, the surplus electricity is curtailed. For periods without solar radiation and with a depleted battery, being an off-grid system, energy/power constraints are like the case I.
- Case IV – The prosumer has a grid-connected photovoltaic system and self-consumption is made. The surplus electricity is sent to the battery storage, which has priority, regarding the injection in the grid. If the battery achieves the maximum state of charge, the surplus electricity is sold to the grid. In periods where the generation isn’t enough to supply the loads, the prosumer consumes electricity from the grid.

The four proposed cases energy flows are summarized in Figure 8.

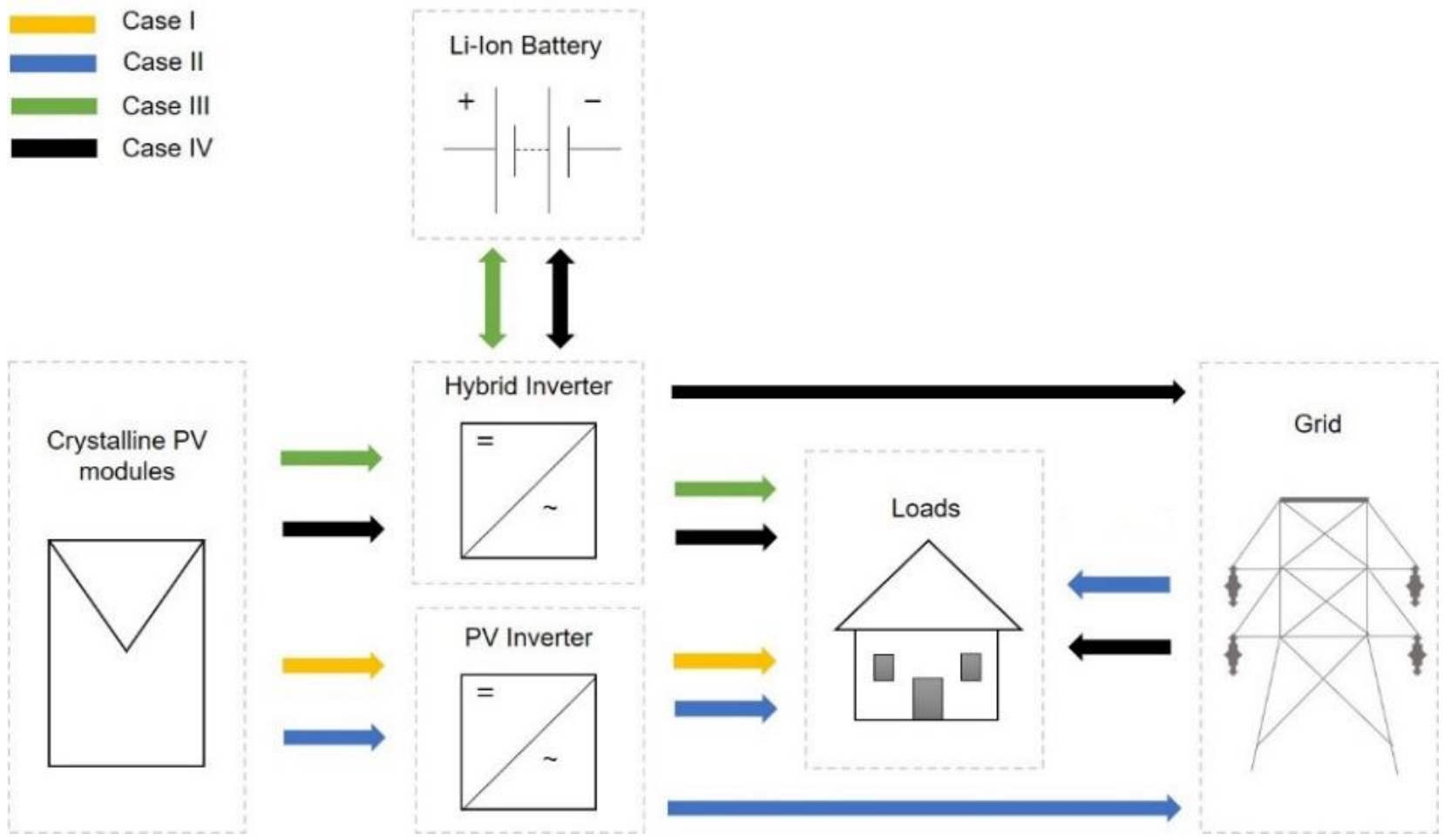

Figure 8 - Energy flows of the proposed case scenarios.

Cases III and IV simulate the use of a PV+battery setup. A simple demonstration of the solar photovoltaic energy flows in these scenarios for 1.50 kW PV power installation in Évora is given in Figure 9.

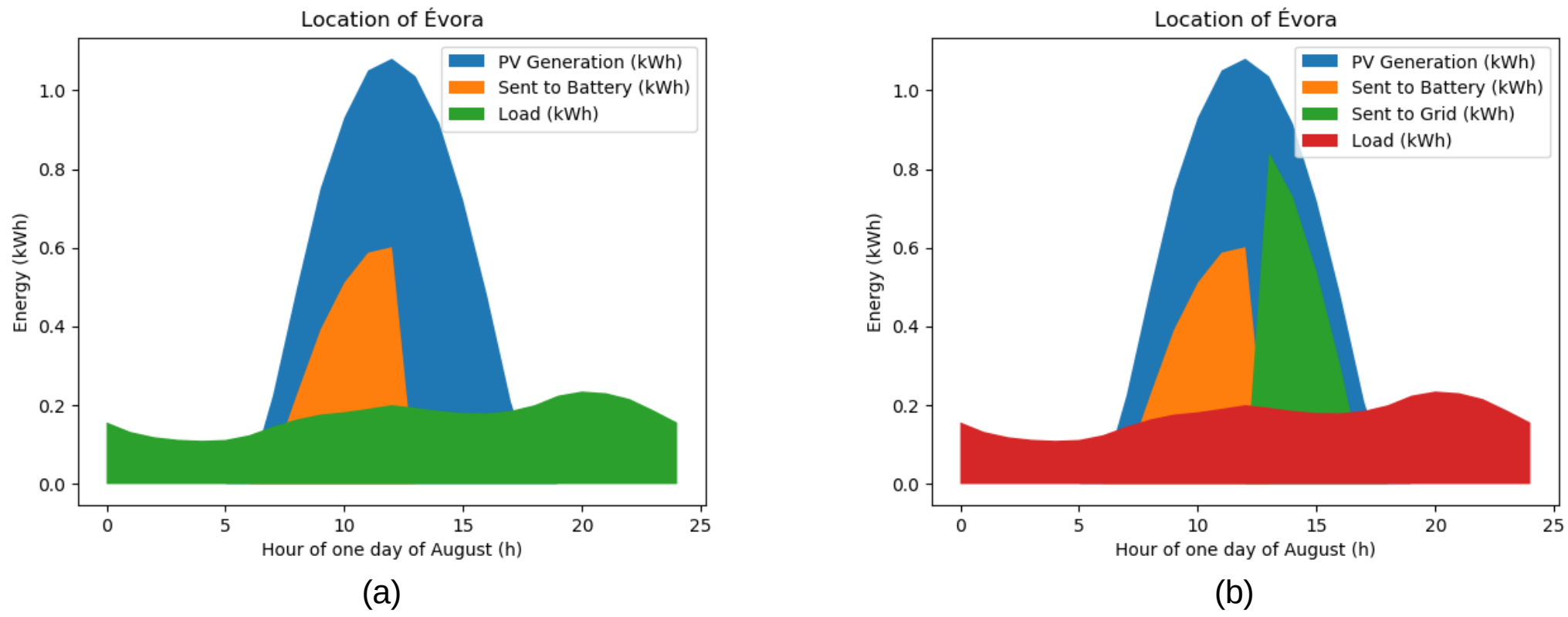

Figure 9 – Example of the Évora 1.50 kW PV installed power with the (a) Case III, energy stored with a 3.3 kWh battery, and (b) Case IV, energy stored to the 3.3 kWh battery and/or exchanged with the grid.

For each configuration, an investment assessment was carried out. Cases II and IV, the grid-connected photovoltaic configurations, consider the contracted power cost in the economic assessment, with the aim of making a real comparison among the studied cases. The assessment is made considering general aspects to all the cases, which are given in Table 7. The component prices are given in Table 8.

Table 7 - General considerations of the case study.

| *Variable* | *Value* |
|---|---|
| *Photovoltaic unit Power (Wp)* | 250 |
| *Photovoltaic Module (€/Wp)* | 0.35 |
| *Power of the Module in year 25 (%)* | 80 |
| *Battery Degradation Capacity (%/year)* | 2.0 |
| *Discount Rate (%)* | 3.0 |
| *Inflation Rate (%)* | 2.5 |

Table 8 - Components prices and considered rates, for the year of 2019.

| | *Case I* | | | | *Case II* | | | | *Case III* | | | | *Case IV* | | | |
|---|---|---|---|---|---|---|---|---|---|---|---|---|---|---|---|---|
| *Identification* | *PV1* | *PV2* | *PV3* | *PV4* | *PV1* | *PV2* | *PV3* | *PV4* | *PV1* | *PV2* | *PV3* | *PV4* | *PV1* | *PV2* | *PV3* | *PV4* |
| *Structures (€)* | 50 | 50 | 200 | 300 | 50 | 50 | 200 | 300 | 50 | 50 | 200 | 300 | 50 | 50 | 200 | 300 |
| *Micro-Inverter or Inverter (€)* | 199 | 324 | 597 | 1393 | 199 | 324 | 597 | 1393 | 1833 | 1833 | 1833 | 1833 | 1833 | 1833 | 1833 | 1833 |
| *Cables and Others (€)* | 50 | 50 | 100 | 100 | 50 | 50 | 100 | 100 | 50 | 50 | 100 | 100 | 50 | 50 | 100 | 100 |
| *Installation (€)* | 100 | 150 | 200 | 300 | 100 | 150 | 200 | 300 | 100 | 150 | 200 | 300 | 100 | 150 | 200 | 300 |
| *Battery (€)* | N/A | | | | N/A | | | | B1 – 1625€;<br>B2 – 4060€;<br>B3 – 5370€. | | | | B1 – 1625€;<br>B2 – 4060€;<br>B3 – 5370€. | | | |
| *Obligations fees by DL-153/2014 (€)* | 0 | 0 | 0 | 0 | 30 | 30 | 30 | 100 | 0 | 0 | 0 | 0 | 30 | 30 | 30 | 170 |

PV1 to PV4 is the selected installed PV power (PV1 = 0.50 kW; PV2 = 0.75 kW; PV3 = 1.50 kW and PV4 = 3.45 kW) and the B1 to B3 represent the several battery capacities. All the component prices were obtained from two Portuguese suppliers and reflect the current real Portuguese market prices [36, 37], verified with the initial market price survey. Potential discounts associated with the purchase of multiple equipment, e.g. several microinverters, were not considered due to the high subjectivity associated with these commercial discounts. An average value for installation cost was also considered, which may have a substantial variability associated with the selected installer but will tend to be more homogeneous (and possibly lower) with the growth and increased competitiveness of the market. The photovoltaic module prices evolved rapidly in the last years, which justifies the use of PV spot market prices, with small approximations reflecting the real costs in Portugal [38]. A remark must be made regarding the bidirectional wattmeter. EDP Distribuição is currently replacing the previous analogue watt meters and deploying new digital versions with bidirectional metering capabilities, in all Portuguese territory, ensuring a 80 % replacement rate until 2020 (European Union directive from 2009). In this way, present analysis ignored the bidirectional counter acquisition costs, obliged by DL 153/2014 whenever its applicable.

## 4. Results

The economic and energetic analysis were made using the interactive computational software MATLAB. In the following tables, the results are shown, with the three/four best results highlighted with a blue coloured cell. Regarding these figures, the initial letter “F” corresponds to the flat tariff, and the letter “B” corresponds to the bi-hourly tariff. In the payback presentation tables, “nan” values correspond to payback periods bigger than 25 years (the considered project lifetime) and are interpreted as uninteresting results for economic analysis.

(A) Economic Analysis

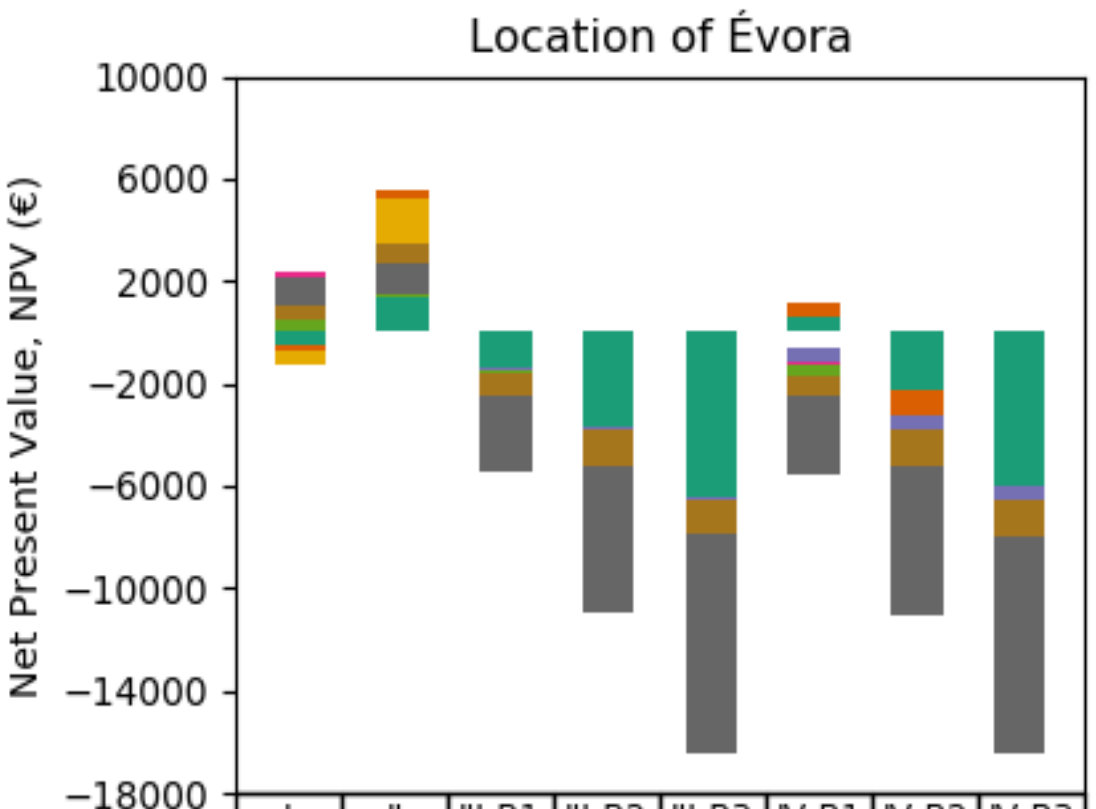

| | I | II | III-B1 | III-B2 | III-B3 | IV-B1 | IV-B2 | IV-B3 |
|---|---|---|---|---|---|---|---|---|
| F_0.50 kW | 1075 | 1230 | -3041 | -5773 | -8504 | -3071 | -5803 | -8534 |
| F_0.75 kW | 1050 | 1504 | -2454 | -5185 | -7917 | -2484 | -5215 | -7947 |
| F_1.50 kW | 485 | 2008 | -1648 | -3790 | -6522 | -1678 | -3820 | -6552 |
| F_3.45 kW | -1217 | 3265 | -2145 | -4702 | -7433 | -2245 | -4802 | -7533 |
| B_0.50 kW | 1228 | 1382 | -1506 | -4237 | -6969 | -1255 | -3987 | -6718 |
| B_0.75 kW | 1181 | 1634 | -1509 | -4240 | -6972 | -1174 | -3906 | -6637 |
| B_1.50 kW | 606 | 2131 | -1391 | -3685 | -6417 | -590 | -3283 | -6015 |
| B_3.45 kW | -1099 | 3393 | -2008 | -4610 | -7341 | 1165 | -2252 | -5988 |

Figure 10 – Results of the NPV economic indicator for the 8th studied configurations, for the location of Évora.

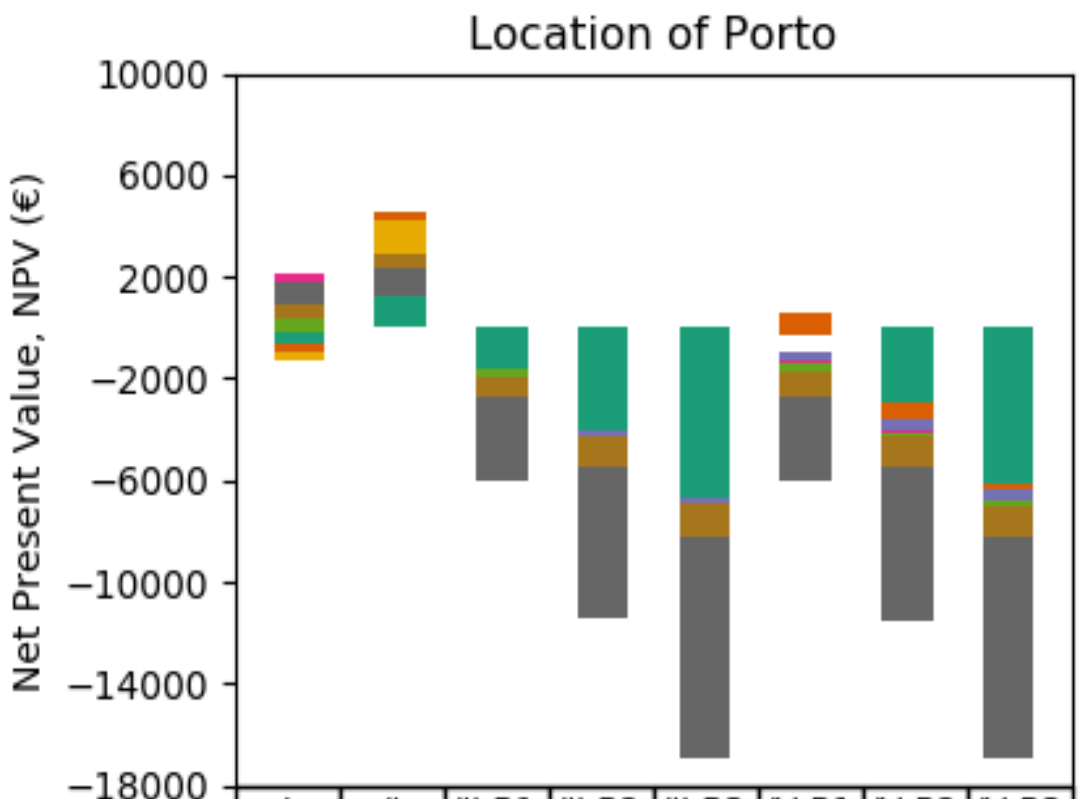

| | I | II | III-B1 | III-B2 | III-B3 | IV-B1 | IV-B2 | IV-B3 |
|---|---|---|---|---|---|---|---|---|
| F_0.50 kW | 939 | 1051 | -3245 | -5977 | -8709 | -3275 | -6007 | -8739 |
| F_0.75 kW | 907 | 1274 | -2734 | -5466 | -8197 | -2764 | -5496 | -8227 |
| F_1.50 kW | 383 | 1665 | -1965 | -4214 | -6946 | -1699 | -4244 | -6976 |
| F_3.45 kW | -1315 | 2568 | -2282 | -4717 | -7449 | -2382 | -4817 | -7549 |
| B_0.50 kW | 1105 | 1217 | -1583 | -4315 | -7046 | -1381 | -4112 | -6844 |
| B_0.75 kW | 1058 | 1427 | -1597 | -4328 | -7060 | -1319 | -4051 | -6782 |
| B_1.50 kW | 514 | 1806 | -1629 | -4002 | -6734 | -967 | -3637 | -6368 |
| B_3.45 kW | -1192 | 2714 | -2113 | -4624 | -7356 | 624 | -2952 | -6171 |

Figure 11 – Economic indicator NPV for the location of Porto, for all the studied configurations.

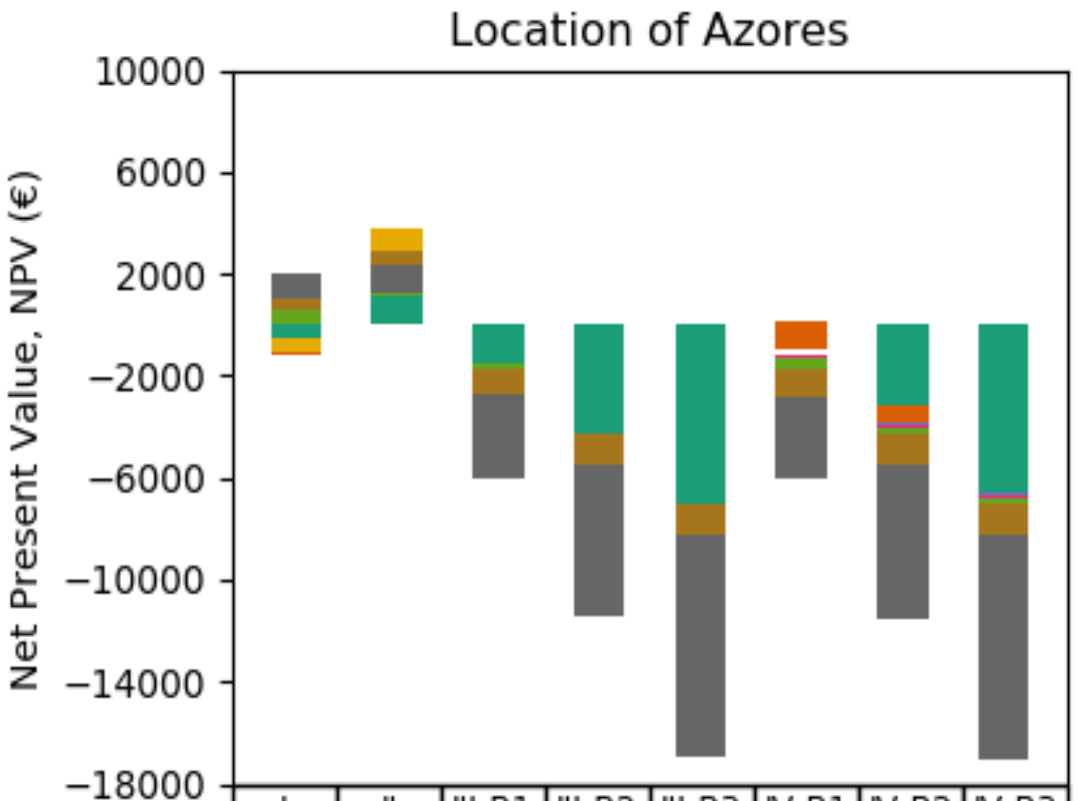

| | I | II | III-B1 | III-B2 | III-B3 | IV-B1 | IV-B2 | IV-B3 |
|---|---|---|---|---|---|---|---|---|
| F_0.50 kW | 1012 | 1063 | -3263 | -5994 | -8726 | -3293 | -6024 | -8756 |
| F_0.75 kW | 1017 | 1277 | -2760 | -5492 | -8223 | -2790 | -5522 | -8253 |
| F_1.50 kW | 552 | 1591 | -1719 | -4288 | -7019 | -1749 | -4318 | -7049 |
| F_3.45 kW | -1063 | 2224 | -1803 | -4255 | -6986 | -1903 | -4355 | -7086 |
| B_0.50 kW | 1047 | 1096 | -1536 | -4268 | -6999 | -1304 | -4036 | -6767 |
| B_0.75 kW | 1010 | 1268 | -1554 | -4285 | -7017 | -1230 | -3962 | -6694 |
| B_1.50 kW | 481 | 1519 | -1676 | -4306 | -7038 | -1132 | -3858 | -6590 |
| B_3.45 kW | -1159 | 2127 | -2090 | -4647 | -7379 | 176 | -3151 | -6697 |

Figure 12 – Location of Azores NPV results, for each of the studied configurations.

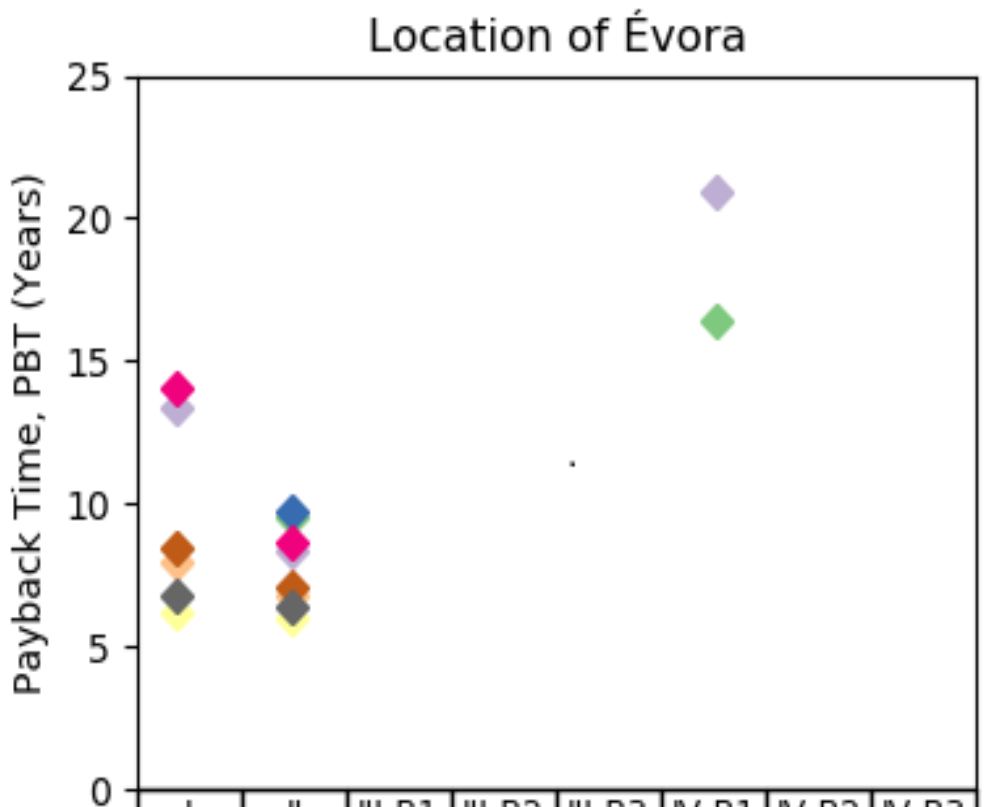

| | I | II | III-B1 | III-B2 | III-B3 | IV-B1 | IV-B2 | IV-B3 |
|---|---|---|---|---|---|---|---|---|
| F_0.50 kW | 6.8 | 6.4 | nan | nan | nan | nan | nan | nan |
| F_0.75 kW | 8.5 | 7.1 | nan | nan | nan | nan | nan | nan |
| F_1.50 kW | 14.1 | 8.6 | nan | nan | nan | nan | nan | nan |
| F_3.45 kW | nan | 9.7 | nan | nan | nan | nan | nan | nan |
| B_0.50 kW | 6.2 | 6.0 | nan | nan | nan | nan | nan | nan |
| B_0.75 kW | 8.0 | 6.7 | nan | nan | nan | nan | nan | nan |
| B_1.50 kW | 13.4 | 8.4 | nan | nan | nan | 21.0 | nan | nan |
| B_3.45 kW | nan | 9.5 | nan | nan | nan | 16.5 | nan | nan |

Figure 13 – Results of the PBT, for the location of Évora, for the configurations studied.

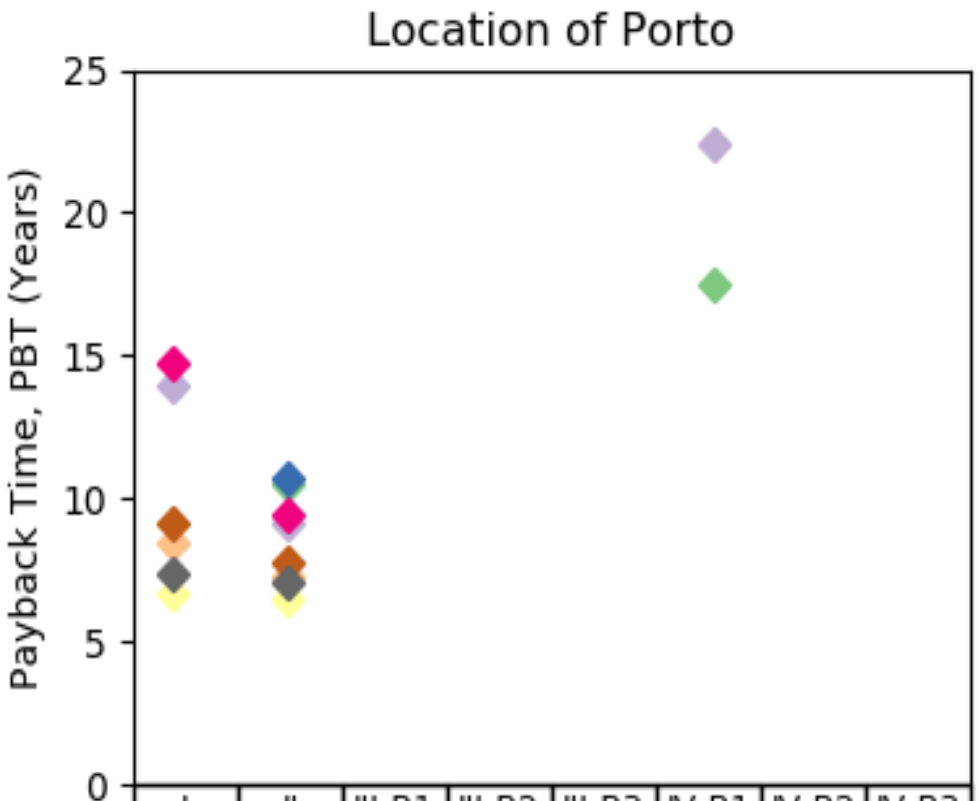

| | I | II | III-B1 | III-B2 | III-B3 | IV-B1 | IV-B2 | IV-B3 |
|---|---|---|---|---|---|---|---|---|
| F_0.50 kW | 7.3 | 7.1 | nan | nan | nan | nan | nan | nan |
| F_0.75 kW | 9.2 | 7.8 | nan | nan | nan | nan | nan | nan |
| F_1.50 kW | 14.7 | 9.5 | nan | nan | nan | nan | nan | nan |
| F_3.45 kW | nan | 10.8 | nan | nan | nan | nan | nan | nan |
| B_0.50 kW | 6.7 | 6.5 | nan | nan | nan | nan | nan | nan |
| B_0.75 kW | 8.5 | 7.3 | nan | nan | nan | nan | nan | nan |
| B_1.50 kW | 13.9 | 9.1 | nan | nan | nan | 22.4 | nan | nan |
| B_3.45 kW | nan | 10.5 | nan | nan | nan | 17.5 | nan | nan |

Figure 14 – Results of the PBT for all the configurations studied, for the location of Porto.

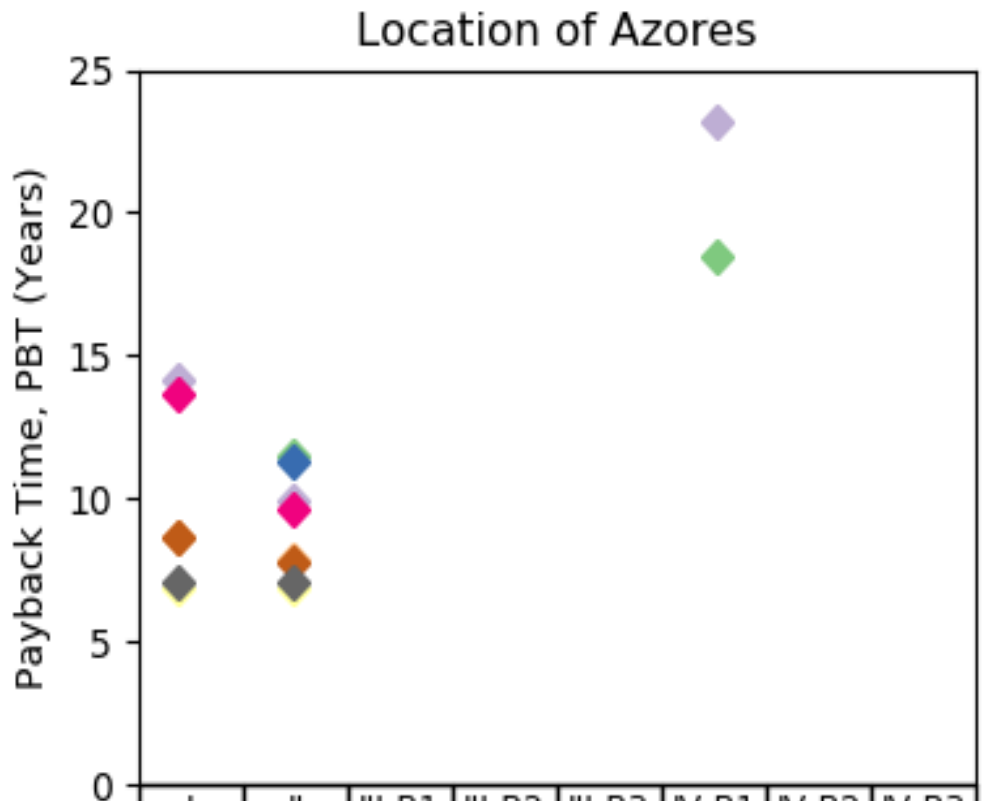

| | I | II | III-B1 | III-B2 | III-B3 | IV-B1 | IV-B2 | IV-B3 |
|---|---|---|---|---|---|---|---|---|
| F_0.50 kW | 7.0 | 7.0 | nan | nan | nan | nan | nan | nan |
| F_0.75 kW | 8.6 | 7.8 | nan | nan | nan | nan | nan | nan |
| F_1.50 kW | 13.7 | 9.7 | nan | nan | nan | nan | nan | nan |
| F_3.45 kW | nan | 11.3 | nan | nan | nan | nan | nan | nan |
| B_0.50 kW | 6.9 | 6.9 | nan | nan | nan | nan | nan | nan |
| B_0.75 kW | 8.7 | 7.8 | nan | nan | nan | nan | nan | nan |
| B_1.50 kW | 14.1 | 9.9 | nan | nan | nan | 23.2 | nan | nan |
| B_3.45 kW | nan | 11.5 | nan | nan | nan | 18.4 | nan | nan |

Figure 15 – Obtained results of the PBT economic indicator for all the studied configurations, for the region of Azores.

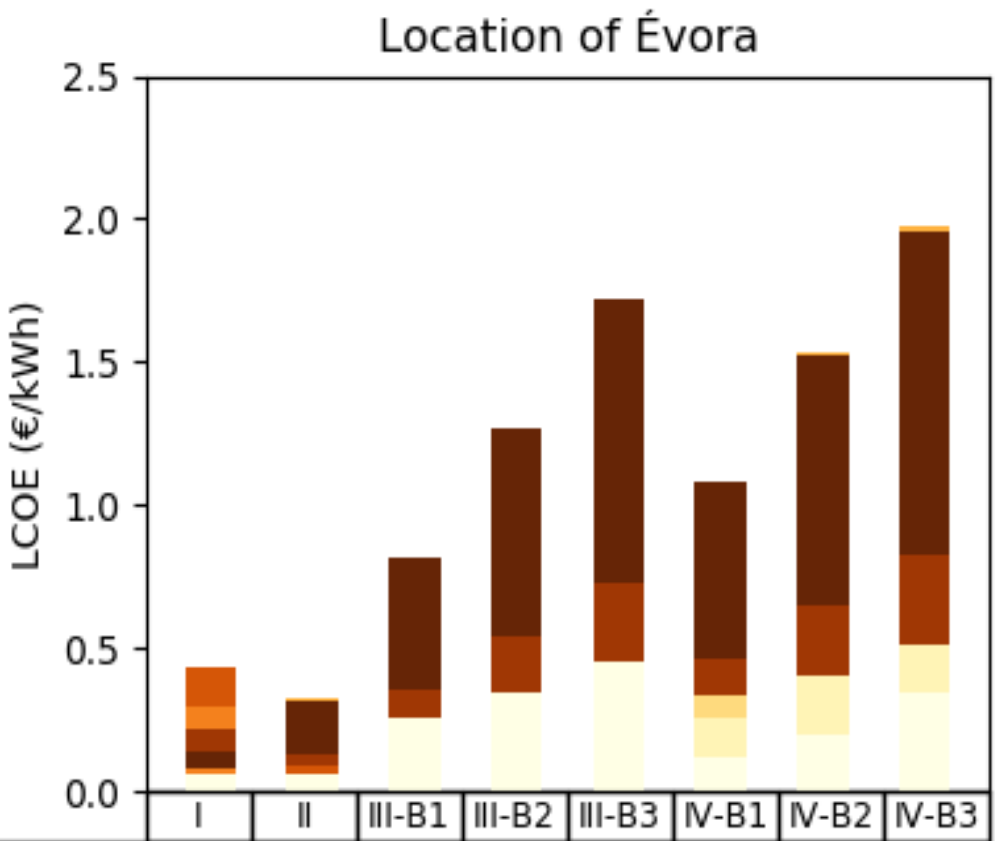

| | I | II | III-B1 | III-B2 | III-B3 | IV-B1 | IV-B2 | IV-B3 |
|---|---|---|---|---|---|---|---|---|
| F_0.50 kW | 0.06 | 0.18 | 0.47 | 0.73 | 0.99 | 0.62 | 0.87 | 1.13 |
| F_0.75 kW | 0.08 | 0.13 | 0.35 | 0.54 | 0.73 | 0.46 | 0.65 | 0.83 |
| F_1.50 kW | 0.14 | 0.09 | 0.26 | 0.34 | 0.45 | 0.33 | 0.40 | 0.51 |
| F_3.45 kW | 0.29 | 0.06 | 0.27 | 0.37 | 0.48 | 0.34 | 0.43 | 0.54 |
| B_0.50 kW | 0.06 | 0.18 | 0.47 | 0.73 | 0.99 | 0.62 | 0.88 | 1.14 |
| B_0.75 kW | 0.08 | 0.14 | 0.35 | 0.54 | 0.73 | 0.46 | 0.65 | 0.84 |
| B_1.50 kW | 0.14 | 0.09 | 0.26 | 0.34 | 0.45 | 0.26 | 0.40 | 0.51 |
| B_3.45 kW | 0.29 | 0.06 | 0.27 | 0.37 | 0.48 | 0.12 | 0.20 | 0.34 |

Figure 16 – Results of the LCOE for the location of Évora, for the studied configurations.

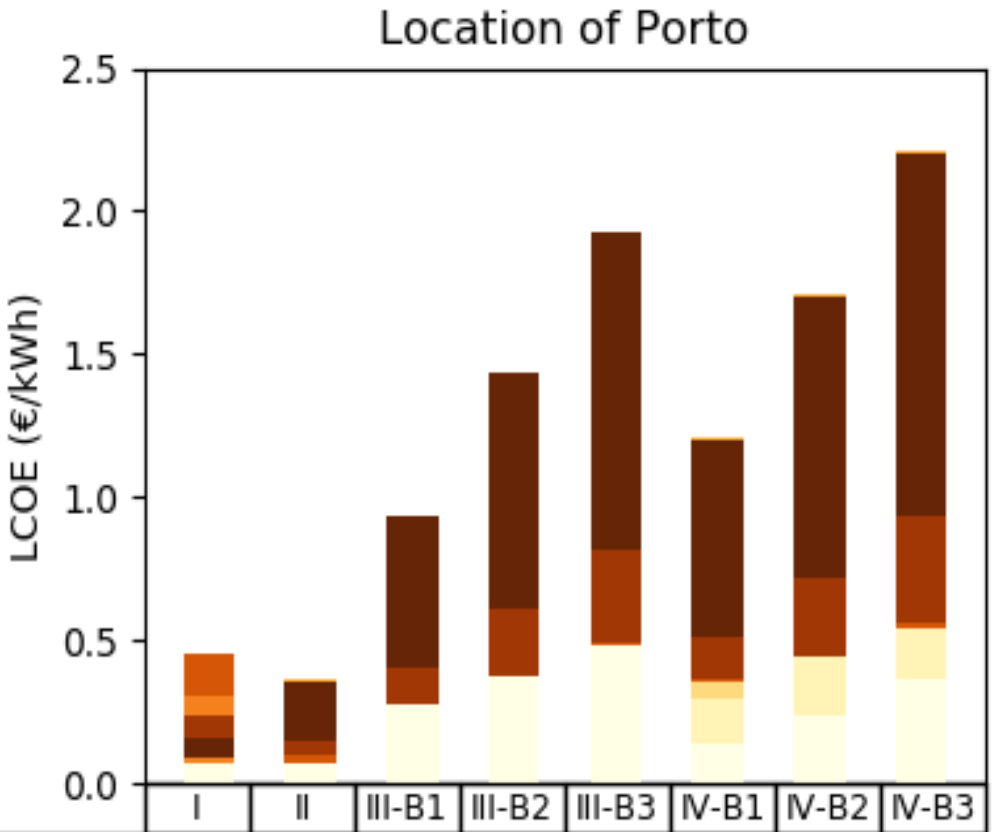

| | I | II | III-B1 | III-B2 | III-B3 | IV-B1 | IV-B2 | IV-B3 |
|---|---|---|---|---|---|---|---|---|
| F_0.50 kW | 0.07 | 0.20 | 0.53 | 0.82 | 1.11 | 0.69 | 0.98 | 1.27 |
| F_0.75 kW | 0.09 | 0.15 | 0.40 | 0.61 | 0.82 | 0.51 | 0.72 | 0.93 |
| F_1.50 kW | 0.15 | 0.10 | 0.28 | 0.37 | 0.49 | 0.36 | 0.44 | 0.56 |
| F_3.45 kW | 0.30 | 0.07 | 0.28 | 0.37 | 0.48 | 0.35 | 0.44 | 0.54 |
| B_0.50 kW | 0.07 | 0.21 | 0.53 | 0.82 | 1.11 | 0.69 | 0.98 | 1.27 |
| B_0.75 kW | 0.09 | 0.15 | 0.40 | 0.61 | 0.82 | 0.52 | 0.73 | 0.94 |
| B_1.50 kW | 0.15 | 0.10 | 0.28 | 0.37 | 0.49 | 0.29 | 0.44 | 0.56 |
| B_3.45 kW | 0.30 | 0.07 | 0.28 | 0.37 | 0.48 | 0.14 | 0.24 | 0.36 |

Figure 17 - Results of the LCOE for the location of Porto, for the studied configurations.

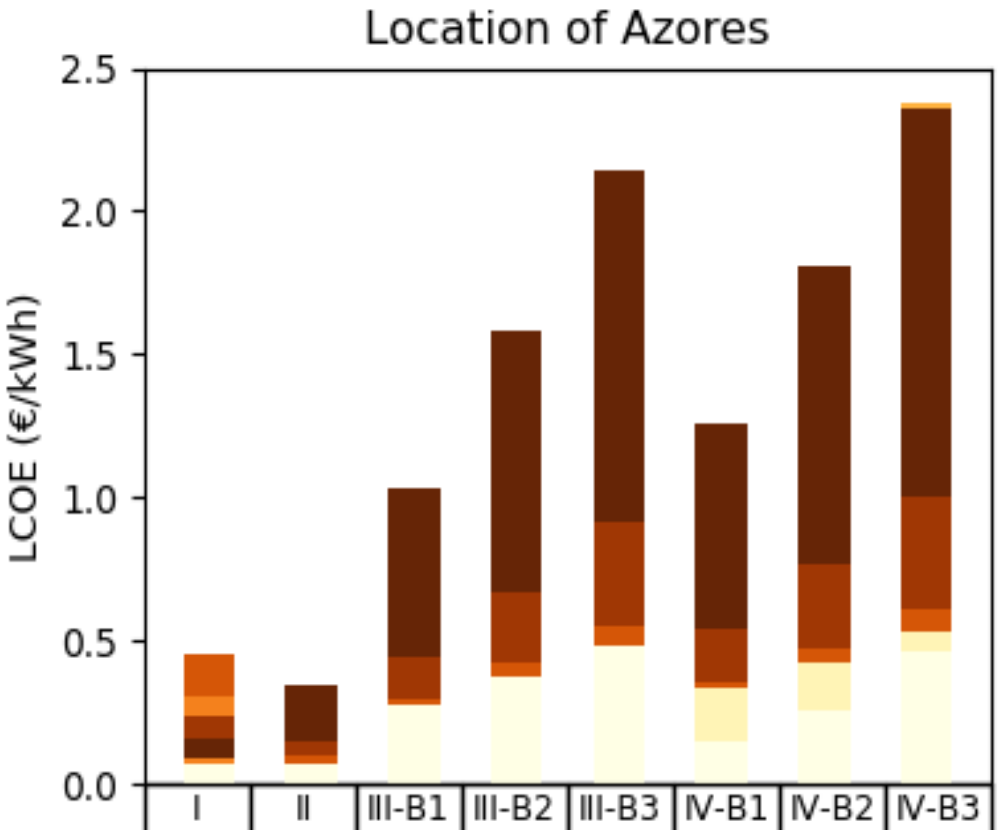

| | I | II | III-B1 | III-B2 | III-B3 | IV-B1 | IV-B2 | IV-B3 |
|---|---|---|---|---|---|---|---|---|
| F_0.50 kW | 0.07 | 0.19 | 0.59 | 0.91 | 1.23 | 0.72 | 1.04 | 1.36 |
| F_0.75 kW | 0.09 | 0.15 | 0.44 | 0.67 | 0.91 | 0.54 | 0.77 | 1.00 |
| F_1.50 kW | 0.15 | 0.10 | 0.29 | 0.42 | 0.55 | 0.35 | 0.47 | 0.61 |
| F_3.45 kW | 0.30 | 0.07 | 0.28 | 0.37 | 0.48 | 0.33 | 0.42 | 0.53 |
| B_0.50 kW | 0.07 | 0.19 | 0.59 | 0.91 | 1.23 | 0.72 | 1.04 | 1.37 |
| B_0.75 kW | 0.09 | 0.15 | 0.44 | 0.67 | 0.91 | 0.54 | 0.77 | 1.01 |
| B_1.50 kW | 0.15 | 0.10 | 0.29 | 0.42 | 0.55 | 0.33 | 0.47 | 0.61 |
| B_3.45 kW | 0.30 | 0.07 | 0.28 | 0.37 | 0.48 | 0.15 | 0.26 | 0.46 |

Figure 18 - Obtained LCOE results for the location of Azores, for all the studied configurations.

## (B) Energy Analysis

The following tables present the energy analysis made for the three chosen locations, for each of the scenarios. For the SMR (Saved Money Rate) indicator the used comparison case was the grid-connected (without battery or PV system), for all the four considered configurations (I to IV).

Table 9 - Energetic analysis for the Évora site.

| **Installed PV Power (kW) / Parameters** | **SCR (Self-consumption Ratio)** | **SSR (Self-supply Ratio** | **BU (Battery Use)** | **SMR (Saved Money Rate)** | | | |
|---|---|---|---|---|---|---|---|
| | **Case I and Case II** | | | **Case I** | | **Case II** | |
| | *Tariff independent* | | | **Flat** | **Bi-hourly** | **Flat** | **Bi-hourly** |
| 0.50 | 0.7601 | 0.3230 | N/A | 0.2439 | 0.2529 | 0.2713 | 0.2787 |
| 0.75 | 0.5797 | 0.3695 | N/A | 0.2790 | 0.2831 | 0.3506 | 0.3510 |
| 1.50 | 0.3238 | 0.4128 | N/A | 0.3117 | 0.3126 | 0.5413 | 0.5309 |
| 3.45 | 0.1450 | 0.4253 | N/A | 0.3211 | 0.3213 | 0.9884 | 0.9559 |
| **Installed PV Power (kW) / Parameters** | **SCR (Self-consumption Ratio)** | **SSR (Self-supply Ratio** | **BU (Battery Use)** | **Saved Money Rate (SMR) (€)** | | | |
| | | | | **Flat** | **Bi-hourly** | **Flat** | **Bi-hourly** |
| | **Case III, Case IV** | **Case III** | **Case III, Case IV** | **Case III** | | **Case IV** | |
| 0.50 B1 | 0.7601 | 0.3230 | 0.0697 | 0.2965 | 0.2814 | 0.2965 | 0.2814 |
| 0.50 B2 | 0.7601 | 0.3230 | 0.0697 | 0.2965 | 0.2814 | 0.2965 | 0.2814 |
| 0.50 B3 | 0.7601 | 0.3230 | 0.0697 | 0.2965 | 0.2814 | 0.2965 | 0.2814 |
| 0.75 B1 | 0.5797 | 0.3695 | 0.1833 | 0.4173 | 0.3961 | 0.4173 | 0.3961 |
| 0.75 B2 | 0.5797 | 0.3695 | 0.1833 | 0.4173 | 0.3961 | 0.4173 | 0.3961 |
| 0.75 B3 | 0.5797 | 0.3695 | 0.1833 | 0.4173 | 0.3961 | 0.4173 | 0.3961 |
| 1.50 B1 | 0.3238 | 0.4128 | 0.4388 | 0.6429 | 0.6101 | 0.7021 | 0.6662 |
| 1.50 B2 | 0.3238 | 0.4128 | 0.5760 | 0.7465 | 0.7084 | 0.7465 | 0.7084 |
| 1.50 B3 | 0.3238 | 0.4128 | 0.5760 | 0.7465 | 0.7084 | 0.7465 | 0.7084 |
| 3.45 B1 | 0.1450 | 0.4253 | 0.5341 | 0.7243 | 0.6874 | 1.1420 | 1.0837 |
| 3.45 B2 | 0.1450 | 0.4253 | 0.5747 | 0.7550 | 0.7164 | 1.0522 | 0.9985 |
| 3.45 B3 | 0.1450 | 0.4253 | 0.5747 | 0.7550 | 0.7164 | 0.9035 | 0.8574 |

Table 10 - Energetic analysis for the location of Porto.

| **Installed PV Power (kW) / Parameters** | **SCR (Self-consumption Ratio)** | **SSR (Self-supply Ratio** | **BU (Battery Use)** | **SMR (Saved Money Rate)** | | | |
|---|---|---|---|---|---|---|---|
| | **Case I and Case II** | | | **Case I** | | **Case II** | |
| | *Tariff independent* | | | **Flat** | **Bi-hourly** | **Flat** | **Bi-hourly** |
| 0.50 | 0.7901 | 0.2964 | N/A | 0.2238 | 0.2356 | 0.2448 | 0.2556 |
| 0.75 | 0.6070 | 0.3416 | N/A | 0.2579 | 0.2659 | 0.3165 | 0.3219 |
| 1.50 | 0.3489 | 0.3927 | N/A | 0.2965 | 0.2998 | 0.4906 | 0.4853 |
| 3.45 | 0.1569 | 0.4061 | N/A | 0.3066 | 0.3082 | 0.8852 | 0.8606 |
| **Installed PV Power (kW) / Parameters** | **SCR (Self-consumption Ratio)** | **SSR (Self-supply Ratio** | **BU (Battery Use)** | **Saved Money Rate (SMR)** | | | |
| | | | | **Flat** | **Bi-hourly** | **Flat** | **Bi-hourly** |
| | **Case III, Case IV** | **Case III** | **Case III, Case IV** | **Case III** | | **Case IV** | |
| 0.50 B1 | 0.7901 | 0.2964 | 0.0538 | 0.2644 | 0.2509 | 0.2644 | 0.2509 |
| 0.50 B2 | 0.7901 | 0.2964 | 0.0538 | 0.2644 | 0.2509 | 0.2644 | 0.2509 |
| 0.50 B3 | 0.7901 | 0.2964 | 0.0538 | 0.2644 | 0.2509 | 0.2644 | 0.2509 |
| 0.75 B1 | 0.6070 | 0.3416 | 0.1512 | 0.3720 | 0.3531 | 0.3720 | 0.3531 |
| 0.75 B2 | 0.6070 | 0.3416 | 0.1512 | 0.3720 | 0.3531 | 0.3720 | 0.3531 |
| 0.75 B3 | 0.6070 | 0.3416 | 0.1512 | 0.3720 | 0.3531 | 0.3720 | 0.3531 |
| 1.50 B1 | 0.3489 | 0.3927 | 0.3889 | 0.5901 | 0.5600 | 0.6338 | 0.6015 |
| 1.50 B2 | 0.3489 | 0.3927 | 0.5011 | 0.6748 | 0.6404 | 0.6748 | 0.6404 |
| 1.50 B3 | 0.3489 | 0.3927 | 0.5011 | 0.6748 | 0.6404 | 0.6748 | 0.6404 |
| 3.45 B1 | 0.1569 | 0.4061 | 0.5249 | 0.7028 | 0.6670 | 1.0599 | 1.0059 |
| 3.45 B2 | 0.1569 | 0.4061 | 0.5939 | 0.7550 | 0.7164 | 0.9545 | 0.9058 |
| 3.45 B3 | 0.1569 | 0.4061 | 0.5939 | 0.7550 | 0.7164 | 0.8825 | 0.8375 |

Table 11 - Energetic analysis for the location of Azores.

| Installed PV Power (kW) / Parameters | SCR (Self-consumption Ratio) | SSR (Self-supply Ratio | BU (Battery Use) | SMR (Saved Money Rate) | | | |
|---|---|---|---|---|---|---|---|
| | Case I and Case II | | | Case I | | Case II | |
| | *Tariff independent* | | | Flat | Bi-hourly | Flat | Bi-hourly |
| 0.50 | 0.8664 | 0.2829 | N/A | 0.2314 | 0.2449 | 0.2432 | 0.2568 |
| 0.75 | 0.6748 | 0.3305 | N/A | 0.2703 | 0.2789 | 0.3126 | 0.3223 |
| 1.50 | 0.3958 | 0.3877 | N/A | 0.3171 | 0.3177 | 0.4731 | 0.4789 |
| 3.45 | 0.1841 | 0.4147 | N/A | 0.3392 | 0.3368 | 0.8229 | 0.8375 |
| **Installed PV Power (kW) / Parameters** | **SCR (Self-consumption Ratio)** | **SSR (Self-supply Ratio** | **BU (Battery Use)** | **Saved Money Rate (SMR) (€)** | | | |
| | | | | **Flat** | **Bi-hourly** | **Flat** | **Bi-hourly** |
| | **Case III, Case IV** | **Case III** | **Case III, Case IV** | Case III | | Case IV | |
| 0.50 B1 | 0.8664 | 0.2829 | 0.0298 | 0.2558 | 0.2648 | 0.2558 | 0.2648 |
| 0.50 B2 | 0.8664 | 0.2829 | 0.0298 | 0.2558 | 0.2648 | 0.2558 | 0.2648 |
| 0.50 B3 | 0.8664 | 0.2829 | 0.0298 | 0.2558 | 0.2648 | 0.2558 | 0.2648 |
| 0.75 B1 | 0.6748 | 0.3305 | 0.1089 | 0.3594 | 0.3722 | 0.3594 | 0.3722 |
| 0.75 B2 | 0.6748 | 0.3305 | 0.1089 | 0.3594 | 0.3722 | 0.3594 | 0.3722 |
| 0.75 B3 | 0.6748 | 0.3305 | 0.1089 | 0.3594 | 0.3722 | 0.3594 | 0.3722 |
| 1.50 B1 | 0.3958 | 0.3877 | 0.3702 | 0.6199 | 0.6419 | 0.6339 | 0.6564 |
| 1.50 B2 | 0.3958 | 0.3877 | 0.4048 | 0.6482 | 0.6712 | 0.6482 | 0.6712 |
| 1.50 B3 | 0.3958 | 0.3877 | 0.4048 | 0.6482 | 0.6712 | 0.6482 | 0.6712 |
| 3.45 B1 | 0.1841 | 0.4147 | 0.5259 | 0.7693 | 0.7966 | 1.0368 | 1.0736 |
| 3.45 B2 | 0.1841 | 0.4147 | 0.5853 | 0.8179 | 0.8469 | 0.9732 | 1.0077 |
| 3.45 B3 | 0.1841 | 0.4147 | 0.5853 | 0.8179 | 0.8469 | 0.8543 | 0.8847 |

## 5. Discussion

As a general comment, cases I and II, which consider the PV-only configurations, are the most profitable. PV+battery configurations are already profitable in very specific conditions, and only with the configuration which has the highest PV power installation (3.45 kW), being slightly better with the bi-hourly tariff. The bi-hourly tariff is the most profitable electric tariff to use in all the cases. Generally, the Azores site configurations are the less profitable and the Évora site are the most profitable, mostly due to the available solar irradiation levels and consequently higher PV power generation. Although energy management strategies are relevant for PV+battery configurations profitability, the geographical location and electric tariff choice are essential factors in the configuration's economic and energetic viability. The 25 years analysis period considers the investment in two batteries units over that time period. This decision was made considering the useful lifecycle of the lithium-ion batteries available indicated in the consulted bibliography and warranty by the battery manufacturers.

In the following, the main three obtained economic indicators are discussed. Regarding NPV, the three locations have similar scenarios, in the sense of a go/no go decision regarding the investment, as can be observed in Figure 10, Figure 11 and Figure 12. Case II is profitable in all configurations regardless of the electricity tariff or location, although in some locations this result is more positive than others. Including a battery is only economically viable when the generated PV energy is the biggest, namely in the IVB1 configuration with 3.45 kW installed PV power, and only for the by-hourly tariff. Case I presents one unviable project, the 3.45 kW size, regardless tariff, since the oversizing of the installation.

Payback time is depicted in Figure 13, Figure 14 and Figure 15. As general remark, the payback time is positive for cases I and II, and mostly negative for cases III and IV. With the considered conditions, case III is unprofitable in all locations. This case doesn't consider the potential additional costs of a RESP connection, generally associated with off-grid connections (since off-grid configurations are usually characterized by being distant from the available grid point of connection), and which would have had impact in the economic indicators in all the studied locations. The best PV+battery scenario for the Évora site is the bi-hourly tariff with 3.45 kW PV installation IVB1, although still considered slightly high - 16 years - compared with a 25 years investment. An interesting aspect is the difference between the obtained results of case IV in Évora, Porto, and in Azores, because of the different (higher) electricity tariff in this last location, and even though associated with the smallest solar resource, presents the worst scenarios, concluding that although having a higher electricity tariff still can't compensate the lowest solar irradiation. Case I average payback is 9.5 years for the location of Évora, 10 years for Porto and 9.8 years for Azores. Case II average payback is 7.8 years for Évora, 8.6 years for Porto and 9.0 years for Azores. This result shows that the grid-connected installations in Portugal have better payback, location independent, due to the increased income of selling the energy surplus to the grid. This means that in average, its 22 % more economic to invest in a grid-connected installation (case II) in Évora, 16 % in Porto and 9 % in Azores.

Considering all sites, in some of the studied situations grid-parity is achieved, observing the obtained LCOE values (Figure 16, Figure 17 and Figure 18). The lowest values, and so best results, are observed in cases I and II, for all PV configurations and all the locations. The use of the batteries presents positive economic interest in case IVB1 for the 3.45 kW PV installation. The lowest PV+battery energy profit occurs in Azores. The most striking economically unviable cases are the case scenarios IIIB3 and IVB3 0.50 kW, due to the low usage of the battery capabilities, and high cost (inexistence of a balanced trade-off). The same comment regarding the case III because of the absence of additional costs associated with the connections with RESP, related to off-grid configurations, which aren't considered in this work.

The IRR obtained results are presented in Table A. 1, Table A. 2 and Table A. 3., and a general comment can be done regarding the high obtained values in the unprofitable configuration's cases. The B/C ratio, present in

Table A. 4, Table A. 5 and Table A. 6, and the higher value, the better project viability. The indicator corroborates the three main discussed economic indicators in this analysis.

The energy analysis made, presented in Table 9, Table 10 and Table 11 allows a more detailed analysis of the generated energy use. The SCR decreases with the increase of the PV installed power, which confirms the existing trade-off between PV generation and effective consumption of this energy. SCR indicator is showing how much of the produced energy is effectively self-consumed and is always dependent of the load diagram and PV generation. In Évora case I, SCR is lower than the one in Porto, and Azores has the highest because the PV generation is the lowest, so the energy ratio is more influenced, compared to the other locations. SSR increases with the increase of the PV installed power, because the biggest the PV generated electricity, the biggest the consumed energy, and considering the energy load constant. The BU indicator helps in the evaluation of need of a battery system, and its value is high when its use is high. In the cases III and IV, the BU indicator confirms that the 0.50 kW and 0.75 kW PV installed power, the PV generation is too low to justify a battery acquisition, so its size is irrelevant in the final gross of energy, in the three locations. With the 1.50 kW and 3.45 kW PV installed power, the use of the battery increases a lot and helps the energy independency. For most of the locations, SMR is smaller for the bi-hourly tariff cases. This indicator compares the energy saved with the current configuration, considering the energy prices at which the electricity is effectively sold, and the electricity bill, in one year, after the payback achievement. The relevance of the introduction of this indicator is mostly to represent the major differences among the use of different electric prices and different electric companies' prices of the contracted power and of energy (EDP commercial and EDA). The fact that Azores has a distinct tariff is well observed in the SMR indicator, because it has the lowest PV generation, but the highest remuneration for the energy makes it have some of the best SMR values. For the case IV, it is noticeable that values above the unity means that one is earning money with the configuration, even though the configuration is already paid. The 23rd article of the DL 153/2014 doesn't establish a limit for the UPAC's injection in RESP, so this configuration is very interesting. The biggest differences between cases III and IV is prominent in the higher PV installed power, as the grid injection remuneration is very low.

# 6. Conclusion

The main aim of this study was the evaluation of the viability of different solar PV configurations in different situations. Four cases were investigated, two cases with PV-only configurations, differing from each other by the injection of the surplus to the grid, and two PV+battery configurations which differ also from the injected surplus, and the inclusion of batteries. The most profitable PV-only configurations for the locations of Évora, Porto and Azores is the case II (0.50 kW PV power with bi-hourly tariff). These are followed in a general way by case I (0.50 kW PV power). The most profitable PV+battery configuration for Évora, Porto and Azores is case IVB1 (3.45 kW PV installed power + 3.3 kWh battery). Although these are very positive results from a PV-only configuration perspective, most of the studied cases of PV+battery aren't profitable, but the scenario shows a very positive future perspective. The bi-hourly tariff presents better results, with the used load profile, which doesn't have a profile with striking load variations. The energy management strategy used in this study was the simplest, but the usage of an intelligent energy management strategy can, by itself, improve the results obtained in this study, particularly considering a multi optimization strategy.

Current average price of the batteries considered in this study is 492 €/kWh, which is still a very high value, and makes the CAPEX of the PV+battery configuration a competitive value, compared with other alternatives. Further decrease of the battery costs which are expected in the following years will be needed to improve the profitability of PV residential applications, although this study is a remark of that beginning. Independent of the technology chosen, battery energy storage has been quickly evolving, with technical improvements being achieved, as for the capacity, performance, efficiency and the response that manufacturers are giving to the market.

All the configurations implemented self-consumption, considered to be the current most adequate context to implement PV solar energy in Portugal in the residential sector, regarding the legislation in force. A revision of the current DL is ongoing due to the evolution that PV technology and batteries have been showing since 2014 (year of the DL prevalence), following the example of different and more profitable residential schemes, as net-metering or community sharing PV generation, from an economic, energetic and social well-being point of view.

## Acknowledgements

The project AGERAR (Ref. 0076_AGERAR_6_E) is co-financed by the European Regional Development Fund (ERDF), within the INTERREG VA Spain-Portugal cooperation programme (POCTEP).

## List of References

## List of Figures

## List of Tables

# Appendices

Table A. 1 - Results of the Internal Rate of Return economic indicator, for the studied configurations cases, for the location of Évora.

| *Location of Évora* | | | | | | | | |
|---|---|---|---|---|---|---|---|---|
| *Electric Tariff* | *Flat* | | | | *Bi-hourly* | | | |
| *Configuration* | *0.5* | *0.75* | *1.5* | *3.45* | *0.5* | *0.75* | *1.5* | *3.45* |
| *I* | 13.0 | 8.41 | -2.42 | -188 | 14.7 | 9.65 | -1.17 | -115 |
| *II* | 14.0 | 12.0 | 8.08 | 5.78 | 15.6 | 13.0 | 8.69 | 6.13 |
| *IIIB1* | -200 | -199 | -194 | -194 | -196 | -195 | -78.0 | -193 |
| *IIIB2* | -202 | -202 | -200 | -200 | -201 | -201 | -200 | -200 |
| *IIIB3* | -203 | -203 | -202 | -201 | -202 | -202 | -202 | -201 |
| *IVB1* | -200 | -199 | -194 | -195 | -106 | -60.1 | -19.0 | -4.66 |
| *IVB2* | -202 | -202 | -200 | -199 | -201 | -200 | -199 | -85.6 |
| *IVB3* | -203 | -203 | -202 | -201 | -202 | -202 | -201 | -200 |

Table A. 2 - Obtained results regarding the IRR economic indicator, for the studied configurations cases, for the location of Porto.

| *Location of Porto* | | | | | | | | |
|---|---|---|---|---|---|---|---|---|
| *Electric Tariff* | *Flat* | | | | *Bi-hourly* | | | |
| *Configuration* | *0.5* | *0.75* | *1.5* | *3.45* | *0.5* | *0.75* | *1.5* | *3.45* |
| *I* | 11.3 | 6.96 | -3.60 | -191 | 13.3 | 8.49 | -2.11 | -187 |
| *II* | 12.0 | 10.1 | 6.27 | 3.72 | 13.9 | 11.4 | 7.04 | 4.17 |
| *IIIB1* | -201 | -200 | -196 | -195 | -196 | -196 | -194 | -194 |
| *IIIB2* | -203 | -202 | -200 | -200 | -201 | -201 | -200 | -200 |
| *IIIB3* | -203 | -203 | -202 | -201 | -202 | -202 | -202 | -201 |
| *IVB1* | -201 | -200 | -194 | -195 | -194 | -120 | -29.5 | -7.12 |
| *IVB2* | -203 | -202 | -200 | -198 | -201 | -201 | -199 | -196 |
| *IVB3* | -203 | -203 | -202 | -201 | -202 | -202 | -201 | -200 |

Table A. 3 - Internal Rate of Return economic indicator results, for the studied configurations cases, for the location of Azores.

| *Location of Azores* | | | | | | | | |
|---|---|---|---|---|---|---|---|---|
| *Electric Tariff* | *Flat* | | | | *Bi-hourly* | | | |
| *Configuration* | *0.5* | *0.75* | *1.5* | *3.45* | *0.5* | *0.75* | *1.5* | *3.45* |
| I | 12.2 | 8.08 | -1.71 | -83.4 | 12.63 | 8.02 | -2.46 | -184 |
| II | 12.2 | 10.1 | 5.86 | 2.58 | 12.56 | 10.04 | 5.44 | 2.24 |
| IIIB1 | -201 | -200 | -195 | -134 | -196 | -195 | -194 | -194 |
| IIIB2 | -203 | -202 | -200 | -199 | -201 | -201 | -200 | -200 |
| IIIB3 | -203 | -203 | -202 | -201 | -202 | -202 | -202 | -201 |
| IVB1 | -201 | -200 | -194 | -191 | -162 | -74.19 | -38.4 | -9.72 |
| IVB2 | -203 | -202 | -200 | -199 | -201 | -200 | -200 | -197 |
| IVB3 | -203 | -203 | -202 | -201 | -202 | -202 | -202 | -201 |

Table A. 4 - Obtained results of the Benefit-to-Cost Ratio economic measure, for each of the studied configurations, for the location of Évora.

| *Location of Évora* | | | | | | | | |
|---|---|---|---|---|---|---|---|---|
| *Electric Tariff* | *Flat* | | | | *Bi-hourly* | | | |
| *Configuration* | *0.5* | *0.75* | *1.5* | *3.45* | *0.5* | *0.75* | *1.5* | *3.45* |
| I | 4.35 | 3.42 | 1.97 | 0.97 | 4.76 | 3.65 | 2.08 | 1.02 |
| II | 1.34 | 1.54 | 1.78 | 2.07 | 1.41 | 1.58 | 1.80 | 2.08 |
| IIIB1 | 0.59 | 0.79 | 1.07 | 1.02 | 1.06 | 1.07 | 1.14 | 1.05 |
| IIIB2 | 0.38 | 0.51 | 0.83 | 0.75 | 0.68 | 0.70 | 0.85 | 0.77 |
| IIIB3 | 0.28 | 0.38 | 0.62 | 0.58 | 0.50 | 0.52 | 0.64 | 0.59 |
| IVB1 | 0.45 | 0.61 | 0.85 | 0.83 | 0.87 | 0.90 | 1.07 | 1.45 |
| IVB2 | 0.32 | 0.43 | 0.70 | 0.64 | 0.61 | 0.64 | 0.78 | 0.99 |
| IVB3 | 0.25 | 0.33 | 0.55 | 0.51 | 0.47 | 0.50 | 0.61 | 0.68 |

Table A. 5- Results of the BC Ratio, for each of the studied configurations, for the location of Porto.

| *Location of Porto* | | | | | | | | |
|---|---|---|---|---|---|---|---|---|
| *Electric Tariff* | *Flat* | | | | *Bi-hourly* | | | |
| *Configuration* | *0.5* | *0.75* | *1.5* | *3.45* | *0.5* | *0.75* | *1.5* | *3.45* |
| I | 3.99 | 3.16 | 1.87 | 0.93 | 4.43 | 3.43 | 2.00 | 0.98 |
| II | 1.21 | 1.39 | 1.61 | 1.86 | 1.29 | 1.45 | 1.64 | 1.88 |
| IIIB1 | 0.53 | 0.71 | 0.99 | 0.99 | 1.03 | 1.04 | 1.08 | 1.03 |
| IIIB2 | 0.34 | 0.46 | 0.75 | 0.75 | 0.67 | 0.68 | 0.79 | 0.76 |
| IIIB3 | 0.25 | 0.34 | 0.57 | 0.58 | 0.49 | 0.50 | 0.59 | 0.59 |
| IVB1 | 0.41 | 0.55 | 0.78 | 0.80 | 0.84 | 0.87 | 0.99 | 1.35 |
| IVB2 | 0.29 | 0.39 | 0.64 | 0.64 | 0.59 | 0.62 | 0.72 | 0.89 |
| IVB3 | 0.22 | 0.30 | 0.50 | 0.51 | 0.46 | 0.48 | 0.57 | 0.66 |

Table A. 6 - Results of the Benefit-to-Cost Ratio, for the location of Azores, for the studied configurations.

| *Location of Azores* | | | | | | | | |
|---|---|---|---|---|---|---|---|---|
| *Electric Tariff* | *Flat* | | | | *Bi-hourly* | | | |
| *Configuration* | *0.5* | *0.75* | *1.5* | *3.45* | *0.5* | *0.75* | *1.5* | *3.45* |
| *I* | 4.19 | 3.36 | 2.03 | 1.04 | 4.28 | 3.34 | 1.96 | 1.00 |
| *II* | 1.48 | 1.64 | 1.78 | 1.89 | 1.48 | 1.61 | 1.72 | 1.84 |
| *IIIB1* | 0.52 | 0.70 | 1.05 | 1.10 | 1.05 | 1.05 | 1.06 | 1.03 |
| *IIIB2* | 0.34 | 0.46 | 0.74 | 0.82 | 0.68 | 0.69 | 0.73 | 0.76 |
| *IIIB3* | 0.25 | 0.34 | 0.56 | 0.64 | 0.50 | 0.51 | 0.55 | 0.59 |
| *IVB1* | 0.43 | 0.57 | 0.87 | 0.93 | 0.91 | 0.94 | 1.01 | 1.33 |
| *IVB2* | 0.30 | 0.40 | 0.65 | 0.73 | 0.63 | 0.66 | 0.72 | 0.90 |
| *IVB3* | 0.23 | 0.31 | 0.50 | 0.58 | 0.48 | 0.50 | 0.56 | 0.62 |